\documentclass[11pt]{article}
\pdfoutput=1
\usepackage{jheppub}
\usepackage{graphicx}
\usepackage{amssymb}
\usepackage{bbm}
\usepackage{mathrsfs}
\usepackage{amsmath,amssymb}
\usepackage{slashed}
\usepackage{hyperref}
\usepackage{caption}
\usepackage{xcolor}
\usepackage{dsfont}
\usepackage{verbatim}
\usepackage{subfig}
\usepackage{amsmath,amssymb,amscd,amsfonts,mathtools}
\usepackage{xifthen}
\usepackage{xcolor}
\definecolor{markgreen}{RGB}{230,243,230}
\definecolor{darkolivegreen}{rgb}{0.33, 0.42, 0.18}
\definecolor{darkpastelgreen}{rgb}{0.01, 0.75, 0.24}
\DeclareMathOperator{\Tr}{Tr}

\usepackage[toc,page]{appendix}
\usepackage{epsfig}
\usepackage{epstopdf}
\usepackage{latexsym}
\usepackage{graphicx}
\usepackage{booktabs}
\usepackage{bbm}
\usepackage{color}
\usepackage{physics}
\usepackage{tensor}
\usepackage{verbatim}
\usepackage{subfig}
\usepackage{tikz}
\usepackage{ifthen}
\usepackage{array}
\usetikzlibrary{matrix}
\usetikzlibrary{decorations.markings,calc,shapes,decorations.pathmorphing,patterns,decorations.pathreplacing,arrows.meta}
\usetikzlibrary{positioning}
\usepackage{hyperref}
\usetikzlibrary{patterns}
\usetikzlibrary{positioning}
\newdimen\mydim
\pdfoutput=1
\makeatletter
\def\@fpheader{\relax}

\newcommand*{\ov}[1]{
  $\m@th\overline{\mbox{#1}}$
}
\newcommand*{\ovA}[1]{
  $\m@th\overline{\mbox{#1}\raisebox{3mm}{}}$
}
\newcommand*{\ovB}[1]{
  $\m@th\overline{\mbox{#1\rule{0pt}{3mm}}}$
}
\newcommand*{\ovC}[1]{
  $\m@th\overline{\mbox{#1\strut}}$
}
\newcommand*{\ovD}[1]{
  $\m@th\overline{\mbox{#1\vphantom{\"A}}}$
}
\newcommand*{\ovE}[1]{
  $\m@th\overline{\raisebox{0pt}[1.2\height]{#1}}$
}
\newcommand*{\ovF}[1]{
  $\m@th\overline{\raisebox{0pt}[\dimexpr\height+1mm\relax]{#1}}$
   Package `calc' can be used as alternative for `\dimexpr'.
}
\newcommand*{\ovG}[1]{
  $\m@th\overline{\raisebox{0pt}[\dimexpr\height+1mm\relax]{#1\vphantom{A}}}$
}
\makeatother
\newboolean{showcomments}
\setboolean{showcomments}{false}
\newcommand\rem[1]{\ifthenelse{\boolean{showcomments}}{{#1}}{}}
\newcommand{\be}{\begin{equation}}
\newcommand{\ee}{\end{equation}}

\usepackage{xifthen}
\newcommand{\dalembert}[1][]{\ifthenelse{\isempty{#1}}{\Box}{#1\Box}}
\usetikzlibrary{decorations.pathmorphing}
\tikzset{snake it/.style={decorate, decoration=snake}}

\title{\Large Algebras, Entanglement Islands, and Observers}
\author{Hao Geng$^{a}$, Yikun Jiang$^{b}$ and Jiuci Xu$^{c}$}
\affiliation{$^{a}$Gravity, Spacetime, and Particle Physics (GRASP) Initiative, Harvard University, 17 Oxford St., Cambridge, MA, 02138, USA.}
\affiliation{$^{b}$Department of Physics, Northeastern University, Boston, MA 02115, USA.}
\affiliation{$^{c}$Department of Physics, University of California, Santa Barbara, CA 93106, USA.}
\emailAdd{haogeng@fas.harvard.edu, phys.yk.jiang@gmail.com, Jiuci\_Xu@ucsb.edu}

\abstract{Some recent work has postulated the existence of an ``observer" for a consistent definition of subregion algebras in gravitational universes. The subregion algebras consist of operators dressed to this ``observer" and are typically Type II von Neumann algebras. Nevertheless, as opposed to standard physical systems, such an ``observer" was postulated to have a Hamiltonian $\hat{H}_{\text{obs}}$ linear in phase space variable. This linear form suggests that the complete dynamics of such an ``observer" should also be controlled by an external system or some underlying degrees of freedom within the system. In this paper, we show that this is exactly the case in the island model. In the island model, we have a gravitational asymptotically anti-de Sitter (AdS) spacetime coupled with a non-gravitational bath, and the diffeomorphism symmetries in the gravitational AdS are spontaneously broken due to the bath coupling. In this setup, the ``observer" is constructed using the Goldstone vector field associated with the spontaneously broken diffeomorphism symmetry, and the external system that also controls the dynamics of the ``observer" is the non-gravitational bath. The basic consistency of the entanglement wedge reconstruction requires operators in the entanglement island to be dressed to this ``observer".  Thus, we establish the result that entanglement islands correspond to emergent Type II$_{\infty}$ von Neumann algebras from the holographic dual perspective. This result relies on assuming the geometric modular flow conjecture. Our study also raises a question for earlier constructions of Type II$_{1}$ von Neumann algebras.}

\begin{document}
\maketitle
\flushbottom
\pagebreak

\section{Introduction}
The algebra of operators localized in a subregion is an elementary concept in quantum field theory that captures experiments performed in the empirical world. A subregion can be thought of as the place where an experiment is performed with the experimental results encoded in the correlators of the operators localized in this subregion. Nevertheless, when gravitational effects are incorporated, the above intuition becomes shaky. This is because gravitational theories are endowed with diffeomorphisms as gauge symmetries.  Thus, the concepts of both subregions and local operators are in apparent tension with the very nature of gravitational theories as they are not invariant under diffeomorphisms in general. This consideration raises the question of how to reconcile our empirical description of the world with quantum gravity.

Subregions can be defined in a dynamical way to be consistent with the nature of gravitational theory. Much like how a specific background geometry is consistent with gravity, a subregion can be defined as the output of a minimization problem \cite{Geng:2020fxl,Balasubramanian:2023dpj}. A natural functional of subregions to minimize is the generalized gravitational entropy
\begin{equation}
    S_{\text{gen}}(\Sigma)=\frac{A(\partial\Sigma)}{4G_{N,0}}+S_{\text{matter}}(\Sigma)\,,\label{eq:Sgen}
\end{equation}
where we only consider bulk Einstein's gravity with bare Newton's constant $G_{N,0}$, $\Sigma$ denotes the subregion and $S_{\text{matter}}(\Sigma)$ is the von Neumann entropy for matter fields, including graviton, in the subregion $\Sigma$. The minimized generalized gravitational entropy in Equ.~(\ref{eq:Sgen}) with respect to $\Sigma$ has been argued to be a well-defined quantity in the UV complete quantum gravity theory \cite{Susskind:1994sm,Kabat:1995jq,Donnelly:2014fua}, as the divergence in the matter entanglement entropy should be removed by the renormalized Newton's constant. Examples of subregions defined in this way are \textit{entanglement wedges} in the AdS/CFT correspondence \cite{Headrick:2014cta,Engelhardt:2014gca,Jafferis:2014lza,Jafferis:2015del,Dong:2016eik,Geng:2021hlu}.

A subregion is only a geometric concept, and the physics associated with it is described by observables localized in it. Typical local observables in quantum field theory are local field operators with appropriate smearing supported within a causally complete subregion. These observables generate the subregion algebra, as we will refer to it in the following discussion.\footnote{The algebra is generated by suitable multiplications and additions of (properly smeared) local operators.}
However, in the quantum theory, local operators are in tension with gravity. This is due to the Heisenberg uncertainty principle which implies that strictly localized operators necessarily have off-diagonal elements in the energy-eigenbasis. Thus, local operators don't commute with the Hamiltonian operator. Moreover, in the standard massless gravity theories, the Gauss' law implies that the energy inside a subregion can be measured outside of it by measuring the gravitational potential. As a result, operators in the standard massless gravity seem to be at best quasilocal, which could have nonzero commutators with other operators spacelike separated from them \cite{Giddings:2018umg}. This consideration is better formulated by saying that local operators in gravity should be appropriately dressed to satisfy the requirement of local diffeomorphism invariance, and much like local charged operators in gauge theory, the dressing is generally implemented using the gravitational Wilson line which goes from the location of the operator insertion to the place where gravity is dynamically decoupled \cite{Donnelly:2015hta,Donnelly:2016auv,Donnelly:2016rvo,Donnelly:2017jcd,Giddings:2018umg,Donnelly:2018nbv, Giddings:2025xym}. The place where gravity is dynamically decoupled is usually taken to be the spatial asymptotic boundary of the gravitational spacetime and the dressed operators are manifestly quasilocal. Thus, the physics in subregions that extend to subregions of the spatial asymptotic boundary can be consistently described using the above quasilocal operators as the dressing can be done using gravitational Wilson lines localized within this type of subregions (see Fig.~\ref{pic:opensubregion}). 

Nevertheless, the above construction would run into trouble for closed subregions whose boundaries are completely within the bulk of the gravitational spacetime (see Fig.~\ref{pic:closedsubregion}). This is because the above quasilocal operators are necessarily dressed by gravitational Wilson lines that would go outside of this type of subregions, so the resulting dressed observables are not sensibly localized within the corresponding subregion. Examples of this type of closed subregions are \textit{entanglement islands} \cite{Geng:2021hlu} which played an important role in the calculation of the Page curve for black hole radiation \cite{Almheiri:2019psf,Penington:2019npb,Geng:2020qvw}. Moreover, not all gravitational spacetimes have spatial asymptotic boundary, so the above construction of quasilocal observables completely breaks down in these cases. In fact, spacetimes without spatial asymptotic boundary are of practical relevance as our current universe is believed to be described by a de Sitter spacetime at large scales, whose spatial sections are compact with no boundary. Hence, the question of how to define local observables in quantum gravity is of both theoretical and practical importance.

Various proposals exist in the literature to address the above question. There are two classes of proposals. The first class is based on the classical intuition that observables can be defined in a relative manner (see \cite{Giddings:2005id} and the references therein). It was suggested that local observables should be defined relative to the features of the background spacetime, for example one could use a supernova as a clock to measure time and some distant distribution of galaxies as a coordinate system to define locations. Nevertheless, local observables constructed in \cite{Giddings:2005id} based on this consideration are in fact not generically locally supported. This is due to the fact that the construction in \cite{Giddings:2005id} didn't intend to extract the specific features of the background that can be used to dress operators locally. The second class of proposals are motivated by the analogy with gauge theory that one could introduce additional degrees of freedom whose diffeomorphism transforms can be used to compensate for the diffeomorphism transforms of the local operators \cite{PhysRevD.21.2185,CRovelli_1991,CRovelli_19912,Chandrasekaran:2022cip,Witten:2023qsv,Witten:2023xze,Goeller:2022rsx}. Nevertheless, as opposed to the usual gauge theory, for which there can be negatively charged objects, which can be used to compensate for the gauge transform of positively charged objects, in gravity there are no negative energy excitations; otherwise the system is unstable. Therefore these additional degrees of freedom were postulated to have uncanonical Hamiltonian which is linear in one phase space variable in some recent work \cite{Chandrasekaran:2022cip,Witten:2023qsv,Witten:2023xze,Jensen:2023yxy,Alexandre:2025rgx}, and it was argued that these additional degrees of freedom should be identified as  ``observers" which are somehow emergent from the underlying theory \cite{Witten:2023qsv,Witten:2023xze}. More specifically, it was proposed in \cite{Chandrasekaran:2022cip,Witten:2023qsv,Witten:2023xze} that one can introduce an additional degree of freedom $\hat{p}$ as the ``observer", whose Hamiltonian is linear in its canonical conjugate $\hat{q}$
\begin{equation}
    \hat{H}_{\text{obs}}=\hat{q}\,,\label{eq:Hup}
\end{equation}
which implies that under the time translation
\begin{equation}
    t\rightarrow t+a\,,\label{eq:nonlinear}
\end{equation}
the degree of freedom $\hat{p}$ transforms as
\begin{equation}
    \hat{p}\rightarrow \hat{p}+a\,.
\end{equation}
Thus, one can define a scalar operator local in time by dressing a local quantum field theory operator $\hat{O}(t)$ to this additional degree of freedom as
\begin{equation}
    \hat{O}^{\text{Phys}}(t)=\hat{O}(t-\hat{p})\,.\label{eq:dressed}
\end{equation}
Other large diffeomorphisms can be fixed in a similar manner by postulating that other components of the stress-energy tensor of the additional degrees of freedom are linear in phase space variables. The resulting subregion algebras are argued to be Type II von Neumann algebras \cite{Chandrasekaran:2022cip,Witten:2023qsv,Witten:2023xze,Jensen:2023yxy,DeVuyst:2024grw,Alexandre:2025rgx,Kolchmeyer:2024fly}. Nevertheless, it is important to understand the physical meaning of these additional degrees of freedom, how they emerge from the underlying gravitational theory, and whether the resulting dressed operators are invariant under all diffeomorphisms, especially the locally supported small diffeomorphisms.\footnote{See \cite{Hoehn:2023axh,DeVuyst:2024pop,Kolchmeyer:2024fly,Witten:2025xuc,Chen:2024rpx,Maldacena:2024spf,Yang:2025lme} for attempts to understand the implications of different forms of the ``observer" in various different contexts.}

Recent progress in \cite{Geng:2024dbl} provides a systematic treatment of the questions discussed above. It is realized in \cite{Geng:2024dbl} that the above two classes of proposals and their respective issues are in fact not independent. The additional degrees of freedom introduced by the second class of proposals should in fact be understood as the collective coordinates of the background. Hence only the backgrounds with strong enough features could provide enough collective coordinates to compensate for all diffeomorphisms. These collective coordinates can be easily constructed by realizing that they are the Goldstone bosons associated with the spontaneously broken diffeomorphisms. Thus, they transform nonlinearly under diffeomorphisms much like Equ.~(\ref{eq:nonlinear}) and can be used to compensate the diffeomorphism transforms of local quantum field theory operators as in Equ.~(\ref{eq:dressed}). Nevertheless, detailed studies of the resulting operator algebras are difficult in the context of \cite{Geng:2024dbl} as a concrete background, including the geometry and the matter profiles, and satisfying all of the above requirements would be complicated, though realistic.

\begin{figure}[h]
    \centering
    \begin{tikzpicture}[scale=0.9]
       \draw[-,very thick,red](0,-2) to (0,2);
       \draw[fill=green, draw=none, fill opacity = 0.1] (0,-2) to (4,-2) to (4,2) to (0,2);
           \draw[-,very thick,red](0,-2) to (0,2);
       \draw[fill=violet, draw=none, fill opacity = 0.2] (0,-2) to (-4,-2) to (-4,2) to (0,2);
       \node at (-2,0)
       {\textcolor{black}{$AdS_{d}$}};
        \node at (2,0)
       {\textcolor{black}{$bath$}};
       \draw [-{Computer Modern Rightarrow[scale=1.25]},thick,decorate,decoration=snake] (-1,-1) -- (1,-1);
       \draw [-{Computer Modern Rightarrow[scale=1.25]},thick,decorate,decoration=snake] (1,1) -- (-1,1);
    \end{tikzpicture}
    \caption{\small{A demonstration of the island model. We couple a gravitational AdS$_{d}$ universe (the violet shaded region) with a non-gravitational bath (the green shaded region) glued on its asymptotic boundary. The coupling is achieved by transparent boundary conditions for matter energy-momentum flux.}}
    \label{pic:islandmodel}
\end{figure}
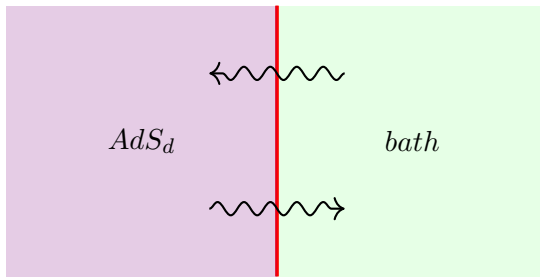

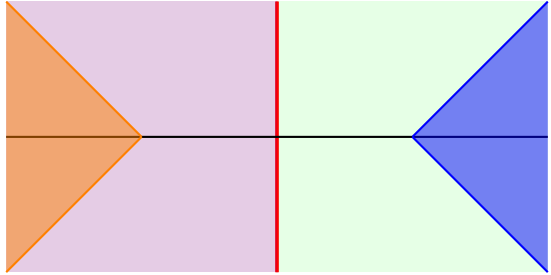
\begin{figure}[h]
    \centering
    \begin{tikzpicture}[scale=0.9]
       \draw[-,very thick,red](0,-2) to (0,2);
       \draw[fill=green, draw=none, fill opacity = 0.1] (0,-2) to (4,-2) to (4,2) to (0,2);
           \draw[-,very thick,red](0,-2) to (0,2);
       \draw[fill=violet, draw=none, fill opacity = 0.2] (0,-2) to (-4,-2) to (-4,2) to (0,2);
       \draw[-,thick,black] (-4,0) to (4,0);
       \draw[-,thick,blue] (2,0) to (4,2);
       \draw[-,thick,blue] (2,0) to (4,-2);
        \draw[fill=blue, draw=none, fill opacity = 0.5] (2,0) to (4,2) to (4,-2) to (2,0);
        \draw[-,thick,orange] (-2,0) to (-4,2);
       \draw[-,thick,orange] (-2,0) to (-4,-2);
        \draw[fill=orange, draw=none, fill opacity = 0.5] (-2,0) to (-4,2) to (-4,-2) to (-2,0);
    \end{tikzpicture}
    \caption{\small{Let's take a Cauchy slice (the black slice) and consider the bath subregion $R$. The domain of dependence of the bath subregion is the blue shaded region. Its entanglement entropy is computed by Equ.~(\ref{eq:SQES}). The orange region in the gravitational bulk denotes domain of dependence of the resulting island $I$.}}
    \label{pic:bathsubregion}
\end{figure}

In this paper, we will study a simpler but more interesting model. Unlike the setup considered in \cite{Geng:2024dbl} where the Goldstone bosons are emergent as the collective coordinates of the symmetry breaking background, we will consider a model which has composite Goldstone bosons due to the underlying interaction with a known external system. In this model, everything is known including the detailed underlying mechanism for the symmetry breaking and the emergence of the Goldstone bosons, yet the background geometries are simple which allows an explicit analysis of the operator algebra. This model is the so-called \textit{island model}, where a gravitational asymptotically anti-de Sitter (AdS) spacetime is coupled to a non-gravitational bath attached to its asymptotic boundary (see Fig.~\ref{pic:islandmodel}). The coupling is achieved by imposing transparent boundary conditions for matter fields in the AdS near its asymptotic boundary. In this model, the AdS bulk geometry is stable against the bath coupling and the effect of the bath is to spontaneously break the AdS diffeomorphisms and induce a mass for the AdS graviton \cite{Porrati:2003sa,Aharony:2006hz} with the detailed mechanism recently understood in \cite{Geng:2023ynk,Geng:2023zhq,Geng:2025rov}. The above mentioned Goldstone bosons can be assembled to a vector field $V^{\mu}(x)$ which transforms nonlinearly under the AdS diffeomorphism
\begin{equation}
   V^{\mu}(x)\rightarrow V^{\mu}(x)-\epsilon^{\mu}(x)\,,\quad\text{while } x^{\mu}\rightarrow x^{\mu}+\sqrt{16\pi G_{N}}\epsilon^{\mu}(x)\,,
\end{equation}
to the leading order in the AdS gravitational coupling constant $G_{N}$. Hereafter, we will call $V^{\mu}(x)$ the \textit{Goldstone vector field} and for simplicity we will only work to the leading nontrivial order of $G_{N}$. Interestingly, this model is particularly suited to study the above questions of subregions and local observables in quantum gravity. Firstly, this is because local operators can be consistently defined in the gravitational AdS due to the existence of the Goldstone vector field, for example a scalar quantum field theory operator $\hat{O}(x)$ can be dressed as
\begin{equation}
    \hat{O}^{\text{Phys}}(x)=\hat{O}(x+\sqrt{16\pi G_{N}}V(x))\,,\label{eq:Vdressed}
\end{equation}
where $x+\sqrt{16\pi G_N}V(x)$ is a short-handed notation for $x^{\mu}+\sqrt{16\pi G_N}V^{\mu}(x)$. Secondly, in this model, closed subregions inside the gravitational AdS naturally exist as entanglement islands of bath subregions. For a given bath subregion $R$ (see Fig.~\ref{pic:bathsubregion} for example), the entanglement island $I$ is the output of the following minimization problem
\begin{equation}
    S(R)=\min\text{Ext}_{I} \Big[\frac{A(\partial I)}{4G_{N}}+S_{\text{matter}}(R\cup I)\Big]\,,\label{eq:SQES}
\end{equation}
with the resulting $S(R)$ as the entanglement entropy of the bath subregion $R$ and the boundary of the island $\partial I$ is called the \textit{quantum extremal surface} \cite{Engelhardt:2014gca}. The formula Equ.~(\ref{eq:SQES}) can be derived by translating the replica path integral in the bath to a gravitational path integral involving the gravitational AdS \cite{Almheiri:2019qdq,Geng:2024xpj}. Thus, the resulting $S(R)$ should be a well-defined quantity in a UV complete quantum gravity theory so as the associated subregion $I$, i.e. the entanglement island. In other words, the algebra of operators localized inside $I$ should be a well-defined algebra in the UV complete quantum gravity theory, in contrast to arbitrary subregions considered in some earlier works.

We will study algebra of operators localized inside the entanglement island of the form Equ.~(\ref{eq:Vdressed}). We first show that, with an appropriate element added, any such algebra is unitarily equivalent to the crossed product of the Type III$_{1}$ algebra by its modular automorphism group. This result relies on assuming the geometric modular flow conjecture \cite{Jensen:2023yxy}. The Type III$_{1}$
algebra is the algebra associated with the entanglement island $I$ for the low energy quantum field theory on a fixed background.\footnote{This point is recently brought up to the study of holography in \cite{Leutheusser:2021frk,Leutheusser:2021qhd,Leutheusser:2022bgi}. See also \cite{Soni:2023fke,Gesteau:2023hbq,Kudler-Flam:2023qfl,Kudler-Flam:2024psh,Gesteau:2024rpt,Kolchmeyer:2024fly,Faulkner:2024gst,Jensen:2024dnl} for other relevant work.} This crossed product algebra is of Type II$_{\infty}$ due to Takesaki's theorem (\cite{10.1007/BF02392041} corollary 97) which implies that the gravitational algebra for operators localized within the entanglement island is also of Type II$_{\infty}$. As opposed to the original Type III$_{1}$ algebra for quantum field theory, the Type II$_{\infty}$ algebra has a trace which can be used to define density matrices and the associated entropy \cite{Witten:2018zxz,Witten:2021unn}. We construct the trace and provide an interpretation of the associated entropy as the generalized gravitational entropy for the entanglement island $I$ which we believe should be a fine-grained entropy that is not UV divergent. As opposed to the earlier work \cite{Chandrasekaran:2022cip,Witten:2023qsv,Witten:2023qsv,Jensen:2023yxy}, our algebra is manifestly diffeomorphism invariant. We also notice that the Hamiltonian of the form Equ.~(\ref{eq:Hup}) is not the canonical Hamiltonian of the ``observer" but merely a term that appears in the linearized Hamiltonian constraint equation due to the St\"{u}ckelberg mass term generated by spontaneous diffeomorphism breaking. Hence, there is no reason to expect that Equ.~(\ref{eq:Hup}) is bounded from below. This observation raises a question for the constructions of Type II$_{1}$ von Neumann algebras from the Type II$_{\infty}$ algebras by projecting Equ.~(\ref{eq:Hup}) for its spectrum to be
bounded from below in \cite{Chandrasekaran:2022cip,Witten:2023qsv,Witten:2023xze,Jensen:2023yxy}. We believe that such a projection needs justifications, which we leave for future work.\footnote{A careful analysis in \cite{Jensen:2024dnl} also suggests that the gravitational algebra of closed subregions in realistic scenarios are of Type II$_{\infty}$. We thank Suvrat Raju for pointing this point.}

This paper is organized as follows. In Sec.~\ref{sec:islandreview} we review relevant background knowledge of entanglement islands and the fact that the graviton is massive in the island model. We will consider a simple setup where the gravitational part of the island model is an empty AdS geometry. The essential lessons easily extend to general cases. In Sec.~\ref{sec:algebra} we study algebra of operators localized inside the entanglement island. We will see that gravitational effects modify the Type III$_{1}$ quantum field theory algebra into a Type II$_{\infty}$ factor. We will see how the ``observer" postulated in earlier works \cite{Chandrasekaran:2022cip,Witten:2023qsv,Witten:2023xze,Jensen:2023yxy} is physically realized and why the resulting algebra is manifestly diffeomorphism invariant. We also construct a trace for the Type II$_{\infty}$ algebra we obtained and study its entropy. In Sec.~\ref{sec:comment} we comment on the issue with the constructions of the Type II$_{1}$ algebra by projecting the Type II$_{\infty}$ algebra in \cite{Chandrasekaran:2022cip,Witten:2023qsv,Witten:2023xze,Jensen:2023yxy}. The paper is concluded with discussions in Sec.~\ref{sec:conclusion}. We provide a concise review of the modular theory and a proof of a theorem we will use in the Appendix.

\begin{figure}[h]
\begin{centering}
\subfloat[An Open Subregion\label{pic:opensubregion}]
{\begin{tikzpicture}[scale=0.9]
\draw[-,very thick,red] (2,0) arc (0:360:2);
 \draw[fill=violet, draw=none, fill opacity = 0.2] (2,0) arc (0:360:2);
 \draw[-,very thick,blue] (1.414,1.414) arc (90:270:2 and 1.414);
 \draw[fill=orange, draw=none, fill opacity = 0.5] (1.414,1.414) arc (90:270:2 and 1.414) arc (-45:45:2);
 \node at (1,0)   {\textcolor{black}{$\cross$}};
 \draw[-, thick,snake it,color=black!50] (1,0) to (2,0);
\end{tikzpicture}
}
\hspace{1.5cm}
\subfloat[A Closed Subregion\label{pic:closedsubregion}]
{\begin{tikzpicture}[scale=0.9]
\draw[-,very thick,red] (2,0) arc (0:360:2);
 \draw[fill=violet, draw=none, fill opacity = 0.2] (2,0) arc (0:360:2);
 \draw[-,very thick, blue] (0.5,0.5) arc (90:450:1 and 0.5);
 \draw[fill=orange, draw=none, fill opacity = 0.5] (0.5,0.5) arc (90:450:1 and 0.5);
 \node at (-0.2,0)   {\textcolor{black}{$\cross$}};
 \draw[-,thick,snake it,color=black!50] (-0.2,0) to (1.414,1.414);
\end{tikzpicture}
}
\caption{\small{A demonstration of the different types of subregions in a gravitational universe. We consider a complete Cauchy slice with the red circle as the asymptotic boundary. The blue curves denote the boundary of the subregions we are considering. The subregions are the orange interior bounded by the blue curve. The cross denotes an operator insertion and the waved line ending on the cross denotes the gravitational Wilson line. \textbf{a)} A subregion whose boundary contains a subregion of the asymptotic boundary. In this case, the gravitational Wilson lines, which we use to dress local operators, can be completely put inside the subregion. \textbf{b)} A subregion within the gravitational universe. In this case, the gravitational Wilson lines necessarily go outside of the subregion.}}
\end{centering}
\end{figure}
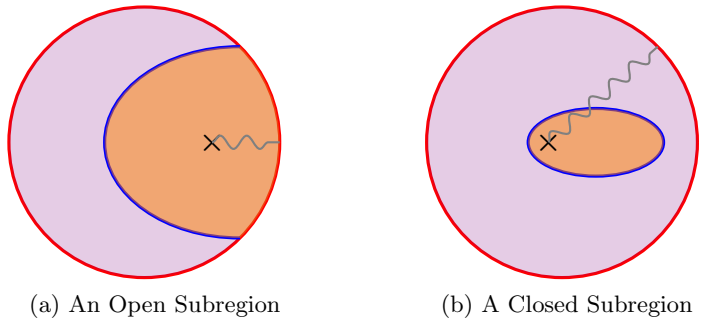

\section{The Physics of Entanglement Islands}\label{sec:islandreview}
In this section, we review the essential physics of entanglement islands relevant to our study. We will first review the fact that the diffeomorphisms are spontaneously broken in the island model and the associated Higgs mechanism.  Then we will review the holographic interpretation of entanglement islands and show how the Higgs mechanism resolves an apparent inconsistency of this interpretation. This review is based on the work \cite{Geng:2025rov} to which we refer the readers for more details.

\subsection{An Explicit Island Model}
We will consider an explicit island model as indicated in Fig.~\ref{pic:islandmodel}. We consider a free massive scalar field  $\phi_{1}(x,z)$ in the gravitational empty AdS$_{d+1}$ spacetime\footnote{In our study, we always consider the weakly gravitating situations and so we treat the whole system perturbatively in $G_{N}$.} as the transparent matter field. For convenience, we model the non-gravitational bath as another AdS$_{d+1}$ spacetime which contains a free massive scalar field $\phi_{2}(x,z)$. We take $\phi_{2}(x,z)$ to have the same mass as $\phi_{1}(x,z)$.\footnote{More specifically, the same $m l_{\text{AdS}}$ with $m$ the mass and $l_{\text{AdS}}$ the AdS length scale.} We take the gravitational and the bath AdS$_{d+1}$'s to be in the Poincar\'{e} patch with metric
\begin{equation}
ds^2=\frac{dz^{2}+\eta_{ij}dx^{i}dx^{j}}{z^{2}}\,,\label{eq:Poincare}
\end{equation}
where $\eta_{ij}$ is the metric for a flat $d$-dimensional Minkowski spacetime and we set all AdS length scales to one. 

The coupling between the two AdS$_{d+1}$'s is achieved by imposing a transparent boundary condition near their asymptotic boundaries such that the energy of $\phi_{1}(x,z)$ in the gravitational AdS$_{d+1}$ can freely leak into $\phi_{2}(x,z)$ in the bath AdS$_{d+1}$.\footnote{In the context of black holes, this enables the black hole to evaporate with the radiation absorbed by the bath \cite{Almheiri:2019psf,Penington:2019npb,Almheiri:2020cfm}.} Under such a leaky boundary condition, the asymptotic behaviors of the two fields $\phi_{1}(x,z)$ and $\phi_{2}(x,z)$ are identified as
\begin{equation}
    \begin{split}
        \phi_{1}(x,z)&=\Big(\alpha(x)z^{\Delta}+\mathcal{O}(z^{\Delta+2})\Big)+\Big(\frac{g}{2\Delta-d}\beta(x)z^{d-\Delta}+\mathcal{O}(z^{d-\Delta+2})\Big)\,,\\\phi_{2}(x,z)&=\Big(\beta(x)z^{d-\Delta}+\mathcal{O}(z^{d-\Delta+2})\Big)+\Big(\frac{g}{d-2\Delta}\alpha(x)z^{\Delta}+\mathcal{O}(z^{\Delta+2})\Big)\,,\label{eq:asympcouple}
    \end{split}
\end{equation}
where $\Delta=\frac{d}{2}+\sqrt{\frac{d^2}{4}+m^{2}}$ with $m^2$ the mass square of the two scalar fields, $\alpha(x)$ and $\beta(x)$ are both dynamical modes and $g$ parametrizes the coupling strength. The above expansion is done in the AdS$_{d+1}$ Poincar\'{e} patch Equ.~(\ref{eq:Poincare}) near the asymptotic boundary $z\rightarrow0$. Besides the leaky matter fields, there exist other fields in the AdS including the graviton and they obey the standard boundary conditions.

The fact that the boundary condition Equ.~(\ref{eq:asympcouple}) couples the gravitational AdS$_{d+1}$ with the bath can be better understood using the AdS/CFT correspondence \cite{Maldacena:1997re,Gubser:1998bc,Witten:1998qj}. The gravitational AdS$_{d+1}$ can be dualized to a CFT$_{d}$ that lives on the asymptotic boundary at $z=0$. In this dual description, the above boundary condition Equ.~(\ref{eq:asympcouple}) indicates that the CFT$_{d}$ is coupled to the non-gravitational bath. Since the bath has the geometry Equ.~(\ref{eq:Poincare}) with the asymptotic boundary also at $z=0$, in this dual description the CFT$_{d}$ lives on the asymptotic boundary of the bath and describes additional boundary degrees of freedom that are coupled to the degrees of freedom of the bath. Using the AdS/CFT correspondence for Equ.~(\ref{eq:asympcouple}), this coupling is in fact achieved by a double-trace deformation for the CFT$_{d}$. The full Hamiltonian for the coupled system is given by
\begin{equation}
    H_{\text{tot}}=H_{\text{CFT$_{d}$}}+g\int d^{d}x O_{1}(x)O_{2}(x)+H_{\text{bath}}\,,\label{eq:coupledH}
\end{equation}
where $O_{1}(x)$ is the CFT$_{d}$ single-trace operator dual to the bulk field $\phi_{1}(x,z)$, $O_{2}(x)$ is defined as the boundary extrapolation of the bath field $\phi_{2}(x,z)$ and $g$ parametrizes the coupling strength between the CFT$_{d}$ and the bath. The extrapolation from $\phi_{2}(x,z)$ to $O_{2}(x)$ is given by
\begin{equation}
    \phi_{2}(x,z)\rightarrow O_{2}(x)z^{d-\Delta}\,,\quad \text{as $z\rightarrow0$}\,.
\end{equation}
The dimension of the operator $O_{1}(x)$ is $\Delta_{1}=\Delta=\frac{d}{2}+\sqrt{\frac{d^2}{4}+m^2}$ and that of $O_{2}(x)$ is $\Delta_{2}=d-\Delta$. Thus, the coupling in Equ.~(\ref{eq:coupledH}) is a marginal deformation for the CFT$_{d}$. As a consistency check, one can work out Equ.~(\ref{eq:asympcouple}) from Equ.~(\ref{eq:coupledH}), using the proposal in \cite{Witten:2001ua} for double-trace deformation in AdS/CFT (see \cite{Geng:2023ynk} for details). From Equ.~(\ref{eq:asympcouple}) or Equ.~(\ref{eq:coupledH}), we obtain the time-ordered Green function of the scalar field $\phi_{1}(x,z)$ in the gravitational AdS$_{d+1}$ as
    \begin{equation}
\begin{split}
\langle \mathbf{T}\phi_{1}(x_{1},z_{1})\phi_{1}(x_{2},z_{2})\rangle&=a_{1}2^{-\Delta_{1}}\frac{\Gamma[\Delta_{1}]}{\pi^{\frac{d}{2}}(2\Delta_{1}-d)\Gamma[\Delta_{1}-\frac{d}{2}]}\frac{_{2}F_{1}[\frac{\Delta_{1}}{2},\frac{\Delta_{1}+1}{2},\Delta_{1}-\frac{d}{2}+1,\frac{1}{Z^{2}}]}{(Z)^{\Delta_{1}}}\\&+a_{2}2^{-\Delta_{2}}\frac{\Gamma[\Delta_{2}]}{\pi^{\frac{d}{2}}(2\Delta_{2}-d)\Gamma[\Delta_{2}-\frac{d}{2}]}\frac{_{2}F_{1}[\frac{\Delta_{2}}{2},\frac{\Delta_{2}+1}{2},\Delta_{2}-\frac{d}{2}+1,\frac{1}{Z^{2}}]}{(Z)^{\Delta_{2}}},\label{eq:2pt}
\end{split}
\end{equation}
where $Z=\frac{z_{1}^{2}+z_{2}^{2}+\eta_{ij}(x_{1}-x_{2})^{i}(x_{1}-x_{2})^{j}}{2z_{1}z_{2}}$ is the invariant distance between the two operator insertions in the gravitational AdS$_{d+1}$,\footnote{For simplicity, the $i\epsilon$-prescription has been ignored.} and we have the relation
\begin{equation}
    a_{2}=\frac{g^{2}}{(2\Delta-d)^{2}}a_{1}\,,
\end{equation}
to the leading order in the coupling constant $g$.

\subsection{The Island Model as a Higgs Phase}
An important feature of the island model is that the matter field correlator Equ.~(\ref{eq:2pt}) induces a mass for the graviton in the gravitational AdS$_{d+1}$ at the one-loop level \cite{Karch:2000ct,Porrati:2003sa,Aharony:2006hz,Geng:2025rov}. This is most straightforwardly seen by the following consideration.

Since we are studying the one-loop effects, we will work to the leading nontrivial order in the Newton's constant $G_{N}$. For convenience, we will denote the AdS$_{d+1}$ coordinates $(x^{i},z)$ as $x^{\mu}$. The gravitational AdS$_{d+1}$ is described by the following path integral
\begin{equation}
Z_{H}=\int D[\psi]D[h_{\mu\nu}]e^{iS_{\text{matter}}[\psi;g^{0}]+iS[h;g^{0}]+iS_{\text{int}}[\psi,h;g^{0}]}\,,\label{eq:Z}
\end{equation}
where the interaction term is given by
\begin{equation}
    S_{\text{int}}[\psi,h;g^{0}]=\frac{\sqrt{16\pi G_{N}}}{2}\int d^{d+1}x\sqrt{-g^{0}(x)}h_{\mu\nu}(x)T^{\mu\nu}_{\text{matter}}(x)\,.
\end{equation}
We have denoted the matter fields collectively as $\psi$, including the transparent matter field $\phi_{1}$ and we used $g^{0}_{\mu\nu}$ to denote the background geometry Equ.~(\ref{eq:Poincare}) with $\sqrt{16\pi G_{N}}h_{\mu\nu}$ denoting the fluctuation of this background metric and $S[h;g^{0}]$ as the kinetic term of the graviton field $h_{\mu\nu}(x)$.\footnote{In linearized gravity, the metric fluctuation is related to the full metric $g_{\mu\nu}$ and the background metric $g^{0}_{\mu\nu}$ as
\begin{equation}
    g_{\mu\nu}=g^{0}_{\mu\nu}+\sqrt{16\pi G_{N}}h_{\mu\nu}\,.
\end{equation}
} To the leading order in $G_{N}$, we have the diffeomorphism transforms
\begin{equation}
\begin{split}
&x^{\mu}\rightarrow x^{\mu}, \quad g^{0}_{\mu\nu}(x)\rightarrow g^{0\prime}_{\mu\nu}(x)=\frac{\partial x'^{\rho}}{\partial x^{\mu}}\frac{\partial x'^{\sigma}}{\partial x^{\nu}}g^{0}_{\rho\sigma}(x')\,,\quad\psi(x)\rightarrow\psi'_{1}(x)=\psi_{1}(x')\,,\\&h_{\mu\nu}(x)\rightarrow h_{\mu\nu}'(x)=\frac{\partial x'^{\rho}}{\partial x^{\mu}}\frac{\partial x'^{\sigma}}{\partial x^{\nu}}h_{\rho\sigma}(x')+\nabla_{\mu}\epsilon_{\nu}(x)+\nabla_{\nu}\epsilon_{\mu}(x)\,,\quad \text{where } x'^{\mu}=x^{\mu}+\sqrt{16\pi G_{N}}\epsilon^{\mu}(x)\,.\label{eq:diffeotr}
\end{split}
\end{equation}
Under the above transforms, the total action $S_{\text{tot}}[\psi,h;g^{0}]=S_{\text{matter}}[\psi;g^{0}]+S[h;g^{0}]+S_{\text{int}}[\psi,h;g^{0}]+S_{\text{bdy}}$ in Equ.~(\ref{eq:Z}) is transformed to
\begin{equation}
    S_{\text{tot}}[\psi,h;g^{0}]\rightarrow  S_{\text{tot}}[\psi,h;g^{0}]+\frac{\sqrt{16\pi G_{N}}}{2}\int d^{d+1}x\sqrt{-g^{0}(x)}\nabla_{\mu}\epsilon_{\nu}(x) T^{\mu\nu}_{\text{matter}}(x)\,,\label{eq:Stottrans}
\end{equation}
where we note that the matter stress-energy tensor $T^{\mu\nu}_{\text{matter}}(x)$ is symmetric and the boundary action $S_{\text{bdy}}$ contains proper terms to ensure a zero on-shell variation. The last term in Equ.~(\ref{eq:Stottrans}) can be written as
\begin{equation}
\begin{split}
    &\int d^{d+1}x\sqrt{-g^{0}(x)}\nabla_{\mu}\epsilon_{\nu}(x) T^{\mu\nu}_{\text{matter}}(x)\,,\\&=\int d^{d+1}x\sqrt{-g^{0}(x)}\nabla_{\mu}\Big(\epsilon_{\nu}(x) T^{\mu\nu}_{\text{matter}}(x)\Big)-\int d^{d+1}x\sqrt{-g^{0}(x)}\epsilon_{\nu}(x) \nabla_{\mu}T^{\mu\nu}_{\text{matter}}(x)\,,\\&=\int d^{d+1}x\sqrt{-g^{0}(x)}\nabla_{\mu}\Big(\epsilon_{\nu}(x) T^{\mu\nu}_{\text{matter}}(x)\Big)\,,
    \end{split}
\end{equation}
where we used the local conservation of the stress-energy tensor inside the gravitational AdS$_{d+1}$. Using Stokes' theorem, the result can be written as a boundary integral with the integrand as the energy-momentum flux of the matter field through the boundary. This flux is not zero due to the transparent matter field $\phi_{1}(x)$ (see \cite{Geng:2023ynk,Geng:2025rov} for details). Thus, the total action is not invariant under all transforms in Equ.~(\ref{eq:diffeotr}).

To have a fully covariant description, we can introduce a compensating vector field $V^{\mu}(x)$ which transforms under the diffeomorphism Equ.~(\ref{eq:diffeotr}) as
\begin{equation}
    V_{\mu}(x)\rightarrow V'_{\mu}(x)=\frac{\partial x^{\nu\prime}}{\partial x^{\mu}} V_{\nu}(x')-\epsilon_{\mu}(x)\,,\label{eq:diffV}
\end{equation}
and integrate it into the path integral Equ.~(\ref{eq:Z}) as
\begin{equation}
    Z_{\text{full}}=\int \frac{D[V^{\mu}]D[\psi]D[h_{\mu\nu}]}{\text{Vol}(G)}e^{iS_{\text{matter}}[\psi;g^{0}]+iS[h;g^{0}]+iS_{\text{int}}[\psi,h;g^{0}]+i\sqrt{16\pi G_{N}}\int d^{d+1}x\sqrt{-g^{0}(x)}\nabla_{\mu}V_{\nu}(x)T^{\mu\nu}_{\text{matter}}(x)}\,,\label{eq:Zfull}
\end{equation}
for which, to ensure consistency, we account for the volume of the gauge group $G$ when introducing the compensating field $V^{\mu}$. This resulting path integral is invariant under the diffeomorphism transforms Equ.~(\ref{eq:diffeotr}) together with Equ.~(\ref{eq:diffV}) to leading order in $G_{N}$. However, we should emphasize that the introduction of the compensating vector field $V^{\mu}(x)$ did not change any physics. This is because the original path integral Equ.~(\ref{eq:Z}) can be realized as a gauge-fixed version of the new path integral Equ.~(\ref{eq:Zfull}) by the gauge fixing condition $V_{\mu}(x)=0$ which has a trivial Feddeev-Popov determinant. This is in contrast to the recent work \cite{Chandrasekaran:2022cip,Witten:2023qsv,Witten:2023xze,Jensen:2023yxy}, where an additional degree of freedom was introduced externally with the physics significantly changed.\footnote{For example, the introduction of this new degree of freedom significantly enlarged the Hilbert space of a gravitational empty de Sitter space \cite{Chandrasekaran:2022cip}.}

We are interested in the one-loop correction of the graviton dynamics by the transparent matter field $\phi_{1}(x)$ in the low-energy regime. We will study this question using the fully covariant description Equ.~(\ref{eq:Zfull}). Thus, we will integrate out the transparent matter field in the path integral Equ.~(\ref{eq:Zfull}) to get an effective action for the graviton and the vector field $V^{\mu}(x)$. Since the full path integral is invariant under the diffeomorphism transforms Equ.~(\ref{eq:diffeotr}) together with Equ.~(\ref{eq:diffV}), the resulting effective action should also be invariant. Hence, the resulting effective action can only be a functional of the following invariant combination
\begin{equation}
    \tilde{h}_{\mu\nu}(x)=h_{\mu\nu}(x)+\nabla_{\mu}V_{\nu}(x)+\nabla_{\nu}V_{\mu}(x)\,.\label{eq:hV}
\end{equation}
A trick to obtain the effective action is to first set the graviton field to be zero, only consider the interaction between the transparent matter field and the vector field $V^{\mu}(x)$ and at the end replace $\nabla_{\mu}V_{\nu}(x)+\nabla_{\nu}V_{\mu}(x)$ by the invariant combination Equ.~(\ref{eq:hV}) \cite{Geng:2023ynk,Geng:2023zhq,Geng:2025rov}. As a result, we have the one-loop quantum effective action
\begin{equation}
    S_{\text{eff}}^{(2)}[h,V]=-\frac{M^{2}}{4}\int d^{d+1}x\sqrt{-g^{0}(x)}\Big(\tilde{h}_{\mu\nu}(x)\tilde{h}^{\mu\nu}(x)-\tilde{h}^{2}(x)\Big)\,,\label{eq:stuckelberg}
\end{equation}
where $M^{2}$ is
\begin{equation}
M^{2}=-G_{N}\frac{2^{4-d}\pi^{\frac{3-d}{2}}}{(d+2)\Gamma(\frac{d+3}{2})}\frac{a_{1}a_{2}\Delta_{1}\Delta_{2}\Gamma[\Delta_{1}]\Gamma[\Delta_{2}]}{\Gamma[\Delta_{1}-\frac{d}{2}]\Gamma[\Delta_{2}-\frac{d}{2}]}\,.\label{eq:mass}
\end{equation}
Interestingly, the quantum effective action Equ.~(\ref{eq:stuckelberg}) is exactly the graviton mass term in the St\"{u}ckelberg form with $V^{\mu}(x)$ as the St\"{u}ckelberg/Goldstone vector boson. This is consistent with the fact that $V^{\mu}(x)$ transforms nonlinearly under the diffeomorphism as in Equ.~(\ref{eq:diffV}). Moreover, if one fixes the unitary gauge $V^{\mu}(x)=0$, the action Equ.~(\ref{eq:mass}) becomes the Fierz-Pauli mass term for graviton which had been known to be the only ghosts free graviton mass term \cite{Hinterbichler:2011tt}.

Moreover, it has recently been understood that the Goldstone vector field $V^{\mu}(x)$ has a nice holographic dual as a composite operator in \cite{Geng:2025rov}. The free equation of motion of the vector field is
\begin{equation}
    \nabla^{2}V^{\nu}(x)+2\nabla_{\mu}\nabla^{\nu}V^{\mu}(x)-2\nabla^{\nu}\nabla_{\mu}V^{\mu}(x)=0\,,
\end{equation}
which is equivalent to
\begin{equation}
    \nabla_{\mu}(\nabla^{\mu}V^{\nu}-\nabla^{\nu}V^{\mu})(x)-2d V^{\nu}(x)=0\,,
\end{equation}
where we used $R_{\mu\nu}=-dg_{\mu\nu}$ for AdS$_{d+1}$. Thus, $V^{\mu}(x)$ is a massive vector field with mass square $m_{V}^{2}=2d$ and we have the near boundary expansion
\begin{equation}
    zV^{\mu}(x,z)\sim z^{d+3}(U^{\mu}(x)+O(z^2))\,,\quad\text{as }z\rightarrow0\,,\label{eq:VUdual}
\end{equation}
where $U^{\mu}(x)$ is the dual operator of the $V^{\mu}(x,z)$\footnote{Note that we have intentionally restored the $z$-coordinate explicit notation.} in the CFT$_{d}$ plus the non-gravitational bath system and it has to be of conformal weight $\Delta_{U}=d+3$. This dual operator is recently identified in \cite{Geng:2025rov} as
\begin{equation}
    U^{\mu}(x)=\langle  \mathcal{O}_{2}(x)\partial^{\mu}\mathcal{O}_{1}(x)\rangle-\mathcal{O}_{2}(x)\partial^{\mu}\mathcal{O}_{1}(x)\,,\label{eq:Udef}
\end{equation}
where $\langle  \mathcal{O}_{2}(x)\partial^{\nu}\mathcal{O}_{1}(x)\rangle$ is the expectation value of the operator $ \mathcal{O}_{2}(x)\partial^{\nu}\mathcal{O}_{1}(x)$. A consistency check for Equ.~(\ref{eq:Udef}) has been performed in \cite{Geng:2025rov} by noticing that its transformation under the CFT$_{d}$ diffeomorphisms are consistent with Equ.~(\ref{eq:VUdual}).\footnote{Moreover, from the dual CFT perspective the fact that the holographic dual of $V^{\mu}(z,x)$ is $U^{\mu}(x)$ is manifest. This is because the operator $O_{2}\partial^{\mu}O_{1}$ controls $\partial_{\mu}T^{\mu\nu}_{\text{CFT}}$ so the CFT stress-energy tensor is in a long multiplet of the conformal symmetry group and the extra components of this representation comparing with the short multiplet for conserved stress-energy tensor are exactly operators $O_{2}\partial^{\mu}O_{1}$\cite{Aharony:2006hz}. The bulk dual of this fact is the Higgs mechanism that the graviton eats the Goldstone vector $V^{\mu}(z,x)$ and becomes massive. } Thus, $U^{\mu}(x)$ is a genuinely composite operator, so as its bulk dual $V^{\mu}(x,z)$ and the island model is in fact in a Higgs phase with composite Goldstone bosons generated from the interaction with the non-gravitational bath. As we will review in the next subsection, this fact is important for the consistency of the holographic interpretation of entanglement islands.

\subsection{The Consistencies of the Holographic Interpretation of Entanglement Islands}

The holographic interpretation of the entanglement islands from the quantum extremal surface formula Equ.~(\ref{eq:SQES}) is that the entanglement island $I$ is part of the entanglement wedge of the bath subregion $R$. This means that the physics in the entanglement island $I$ is fully encoded in the bath subregion $R$ by the entanglement wedge reconstruction \cite{Headrick:2014cta,Engelhardt:2014gca,Jafferis:2014lza,Jafferis:2015del,Dong:2016eik,Penington:2019npb,Almheiri:2019psf}. 

Interestingly, it was realized in \cite{Geng:2021hlu,Geng:2023zhq} that there is an important question for the holographic interpretation of islands. Intuitively, this question can be seen as follows. In the context of the island model, the gravitational Gauss' law for the standard massless gravity would imply that the Hamiltonian of the gravitational AdS$_{d+1}$ is a boundary term which is thus fully accessible near its asymptotic boundary. Since the entanglement island is a closed subregion within the gravitational AdS$_{d+1}$, the above consideration would imply that operators inside the island would not commute with the boundary Hamiltonian. Since the boundary Hamiltonian is an operator outside of the entanglement island, this result contradicts the above holographic interpretation of entanglement islands. This contradiction manifests itself, because $\bar{I}$, the complementary of the entanglement island in the gravitational AdS$_{d+1}$, is part of the entanglement wedge of $\bar{R}$, i.e the complementary region of the subregion $R$ in the non-gravitational bath, and as a result operators inside $I$ and operators in $\bar{I}$ should exactly commute with each other as operators in $R$ and operators in $\bar{R}$. A similar consideration would also suggest that there exist no nontrivial operators in a closed universe, for example de Sitter space, and so the associated quantum gravitational Hilbert space is of dimension one. This question and its answer brought up in \cite{Geng:2021hlu} is the motivation for recent studies of algebras and observers in closed universes \cite{Chandrasekaran:2022cip,Witten:2023qsv}.

In the context of islands, the above intuition can be formulated in a precise way in general relativity using the ADM formalism \cite{Arnowitt:1962hi}. One starts with the ADM decomposition of the metric \cite{Arnowitt:1962hi} on a ($d+1$)-dimensional spacetime
\begin{equation}
ds^{2}=-N^{2} dt^{2}+g_{ij}(dx^{i}+N^{i}dt)(dx^{j}+N^{j}dt)\,,\label{eq:ADM}
\end{equation}
where $N$ is called the \textit{lapse function}, the vector $N^{i}$ is called the \textit{shift vector}. Plugging the metric Equ.~(\ref{eq:ADM}) into the Einstein-Hilbert action, one can see that the canonical momenta associated the lapse function $N$ and shift vector $N^{i}$ are zero, i.e. we have
\begin{equation}
\Pi=\frac{1}{\sqrt{g}}\frac{\delta S}{\delta \dot{N}}=0\,,\quad \Pi_{i}=\frac{1}{\sqrt{g}}\frac{\delta S}{\delta \dot{N^{i}}}=0\,.\label{eq:primary}
\end{equation}
These are dynamical constraints and are called the \textit{primary constraints} \cite{dirac2001quantum}. The primary constraints have to be preserved under the time evolution. This time evolution can be understood if one goes to the Hamiltonian formalism \cite{Arnowitt:1962hi} and the results, from requiring Equ.~(\ref{eq:primary}) to be preserved under the time evolution, are the following \textit{secondary constraints}
\begin{equation}
\mathcal{H}=0\,,\quad \mathcal{H}_{i}=0\,,\label{eq:secondcons}
\end{equation}
where
\begin{equation}
\begin{split}
  \mathcal{H}&=16\pi G_{N}\Big(\Pi_{ij}\Pi^{ij}-\frac{1}{d-1}(\Pi^{i}_{i})^{2}\Big)-\frac{1}{16\pi G_{N}}(R[g]-2\Lambda)+\mathcal{H}_{\text{matter}}\,,\\
\mathcal{H}_{i}&=-2g_{ij}D_{k}\Pi^{jk}+\mathcal{H}_{i,\text{matter}}\,.\label{eq:constraints}
  \end{split}
\end{equation}
We have $\mathcal{H}_{\text{matter}}$ as the Hamiltonian density of the matter fields, $\mathcal{H}_{i,\text{matter}}$ as the matter fields' momentum density and $\Pi^{ij}$ as the canonical momentum of the spatial metric $g_{ij}$
\begin{equation}
\Pi^{ij}=\frac{1}{\sqrt{g}}\frac{\delta S}{\delta \dot{g}_{ij}}=-\frac{1}{16\pi G_{N}}\Big(K^{ij}-g^{ij}K\Big)\,,
\end{equation}
for which $K_{ij}$ are the extrinsic curvature of the constant-$t$ hypersurfaces. The extrinsic curvature can be calculated using the metric Equ.~(\ref{eq:ADM}) as
\begin{equation}
K_{ij}=\frac{1}{2N}(-\dot{g}_{ij}+D_{j}N_{i}+D_{i}N_{j})\,,\label{eq:K}
\end{equation}
where $D_{i}$ is the torsion-free and metric-compatible covariant derivative of the spatial metric $g_{ij}$. In quantum theory, secondary constraints Equ.~(\ref{eq:secondcons}) constrain the physical states and observables. Such constraints can be implemented once we promote $\mathcal{H}$ and $\mathcal{H}_{i}$ to operators $\hat{\mathcal{H}}$ and $\hat{\mathcal{H}}_{i}$. We are interested in exploiting these constraints for physical observables. Gauge invariant observables have to commute with the operators $\hat{\mathcal{H}}$ and $\hat{\mathcal{H}}_{i}$. We will focus on the Hamiltonian constraint $\mathcal{H}=0$, as the momentum constraints $\mathcal{H}_{i}=0$ are easily satisfied, since they basically require that the observables are invariant under the spatial diffeomorphisms.

We are interested in the effect of the Hamiltonian constraint to the leading nontrivial order in $G_{N}$, and we are working around the empty AdS$_{d+1}$ background Equ.~(\ref{eq:Poincare}). Thus we will treat the metric $g_{ij}$ perturbatively and expand it around the background Equ.~(\ref{eq:Poincare}) as
\begin{equation}
g_{ij}=g_{ij}^{0}+\sqrt{16\pi G_{N}}h_{ij}\,.
\end{equation}
In this analysis, at the leading order in $G_{N}$, we have the zeroth order Hamiltonian constraint
\begin{equation}
\mathcal{H}^{(0)}=-\frac{1}{16\pi G_{N}}(R[g^{0}]-2\Lambda)=0\,,
\end{equation}
which is satisfied by the background geometry Equ.~(\ref{eq:Poincare}) with the AdS cosmological constant $\Lambda=-\frac{d(d-1)}{2}$. To the first order in $G_{N}$, we have
\begin{equation}
\mathcal{H}^{(1)}=-\frac{1}{\sqrt{16\pi G_{N}}}\Big[(d-1)h+\hat{\nabla}_{i}\hat{\nabla}_{j}h^{ij}-\hat{\nabla}^{2}h\Big]+\mathcal{H}_{\text{matter}}+\mathcal{H}_{\text{graviton}}=0\,,\label{eq:core}
\end{equation}
in which $h$ is the trace of $h_{ij}$ under the background metric $g^{0}_{ij}$ and $\hat{\nabla}_{i}$ is the torsion free and metric compatible covariant derivative with respect to the background spatial metric. This is the constraint at the first nontrivial order in $G_{N}$, which already distinguishes quantum field theory in a fixed background with quantum gravity, and we will study this first order constraint. In this first order constraint equation, we have $\mathcal{H}_{\text{matter}}$ as the Hamiltonian density of matter fields in the background geometry Equ.~(\ref{eq:Poincare}), and $\mathcal{H}_{\text{graviton}}$ as the free graviton Hamiltonian in the same background geometry. For the sake of simplicity, we will not distinguish them and call them collectively as $\mathcal{H}_{\text{QFT}}$ hereafter. From Equ.~(\ref{eq:core}), we can get an integrated constraint equation\footnote{We are expanding around the AdS Poincar\'{e} patch. The lapse function and shift vector are given by
\begin{equation}
    N=\frac{1}{z}\,,\quad N_{i}=0\,.
\end{equation}} 
\begin{equation}
\begin{split}
\sqrt{16\pi G_{N}}H_{\text{QFT}}&=\sqrt{16\pi G_{N}}\int d^{d}x
\sqrt{g^{0}}N\mathcal{H}_{\text{QFT}}=\int d^{d}x
\sqrt{g^{0}}N\Big[(d-1)h+\hat{\nabla}_{i}\hat{\nabla}_{j}h^{ij}-\hat{\nabla}^{2}h\Big]\,,\\&=\int d^{d}x
\sqrt{g^{0}}\hat{\nabla}_{i}\Big[\frac{\hat{\nabla}_{j}h^{ij}-\hat{\nabla}^{i}h}{z}+\frac{1}{z^{2}}h^{zi}-\frac{1}{z^{2}}h\delta^{i}_{z}\Big]\,,\\&=\int d^{d}x\sqrt{g^{0}}\hat{\nabla}_{i}\Bigg[N\Big[\hat{\nabla}_{j}h^{ij}-\hat{\nabla}^{i}h+\frac{1}{z} (h^{zi}-h\delta^{i}_{z} )\Big]\Bigg]\,,\\&\equiv H_{\partial}\,,\label{eq:HADM}
\end{split}
\end{equation}
Imposing the Dirichlet boundary condition for the graviton field near the asymptotic boundary, we have 
\begin{equation}
    H_{\partial}=\sqrt{16\pi G_{N}}H_{\text{ADM}}\,,
\end{equation}
where $H_{\text{ADM}}$ is the ADM Hamiltonian for asymptotically AdS spacetimes \cite{Arnowitt:1962hi,Hawking:1995fd,Giddings:2018umg}. Thus, the above constraint can be imposed on physical operators as
\begin{equation}
    [\sqrt{16\pi G_{N}}\hat{H}_{\text{QFT}}-\hat{H}_{\partial},\hat{O}^{\text{Phys}}(x)]=0\,,\label{eq:commutator}
\end{equation}
which implies that
\begin{equation}
[\hat{H}_{\partial},\hat{O}^{\text{Phys}}(x)]=\sqrt{16\pi G_{N}}[\hat{H}_{\text{QFT}},\hat{O}^{\text{Phys}}(x)]=-i\sqrt{16\pi G_{N}} \frac{\partial}{\partial t}\hat{O}^{\text{Phys}}(x)\,.\label{eq:constraintO}
\end{equation}
This result is in fact true to all orders in $G_{N}$ \cite{Chowdhury:2021nxw}. Explicit solutions of Equ.~(\ref{eq:constraintO}) for $\hat{O}^{\text{Phys}}(x)$ can be constructed from quantum field theory operators $\hat{O}(x)$ using gravitational Wilson lines \cite{Donnelly:2018nbv,Giddings:2018umg}. This is a concrete formulation of the apparent tension between entanglement islands and the gravitational Gauss' law we mentioned above. While this result may appear puzzling, it aligns with expectations from standard massless gravity theories, where the conventional notion of holography is believed to hold. In such cases, information in the gravitational bulk is expected to be fully encoded in the asymptotic boundary, making it subtle to reconcile with scenarios like islands where the same information also appears to be accessible from an external system. Any attempt to reconcile the situation must take into account the implications of the no-cloning theorem. This highlights the need for careful treatments of information localization in gravitational systems.

In fact, the above consistency question has a nice answer in the island model because of the nonzero graviton mass \cite{Geng:2021hlu,Geng:2023zhq,Geng:2025rov}. More precisely, the additional graviton mass term Equ.~(\ref{eq:stuckelberg}) from the quantum correction of the leaky matter field will modify the Hamiltonian constraint Equ.~(\ref{eq:constraints}). The modification can be easily extracted by noticing that the Hamiltonian constraint is just the $00$-component of the matter coupled Einstein's equation in the ADM decomposition Equ.~(\ref{eq:ADM}).\footnote{The momentum constraints $\mathcal{H}_i=0$ are the $0i$-components of the matter coupled Einstein's equation.} To the first nontrivial order in $G_{N}$ the modified constraint is
\begin{equation}
\begin{split}
    \mathcal{H}^{(1)}=-\frac{1}{\sqrt{16\pi G_{N}}}\Big[(d-1)h+\hat{\nabla}_{i}\hat{\nabla}_{j}h^{ij}-\hat{\nabla}^{2}h\Big]+\mathcal{H}_{\text{QFT}}-\frac{M^{2}}{\sqrt{16\pi G_{N}}}\tilde{h}_{i}^{i}(x)=0\,,\label{eq:stuckelbergH}
    \end{split}
\end{equation}
where $\tilde{h}_{i}^{i}$ is given by the invariant combination
\begin{equation}
    \tilde{h}_{i}^{i}=h_{i}^{i}+2\nabla_{i}V^{i}\,,\label{eq:mHamiltonian}
\end{equation}
where $\nabla_{\mu}$ is the torsion free and metric compatible covariant derivative with the full background metric Equ.~(\ref{eq:Poincare}) instead of just the spatial components. We notice that the additional term in Equ.~(\ref{eq:stuckelbergH}) is the conjugate momentum of the $0$-th component of the St\"{u}ckelberg vector field $V^{\mu}(x)$, which can be obtained from Equ.~(\ref{eq:stuckelberg}) as 
\begin{equation}
    \hat{\pi}_{V^{0}}(x)=M^{2}\Big(h_{i}^{i}+2\nabla_{i}V^{i}\Big)(x)\,,
\end{equation}
for which there is the equal-time commutator
\begin{equation}
    [\hat{\pi}_{V^{0}}(\vec{x},t),V^{0}(\vec{y},t)]=-i\frac{1}{N\sqrt{g^{0}(x)}}\delta^{d}(\vec{x}-\vec{y})\,.\label{eq:canonicalV0}
\end{equation}
As a result, the modified integrated constraint can be written as\footnote{Again, we emphasize that we have used the convention that matter fields include perturbative graviton.}
\begin{equation}
    \sqrt{16\pi G_{N}}\hat{H}_{\text{QFT}}-\hat{\Pi}_{V^{0}}-\hat{H}_{\partial}=0\,,\label{eq:constraintMass}
\end{equation}
which constrains physical operators and we have defined the integrated canonical momentum
\begin{equation}
    \hat{\Pi}_{V^{0}}=\int d^{d}\vec{x} N\sqrt{g^{0}}\hat{\pi}_{V^{0}}(x,t)\,.
\end{equation}

Operators obeying the constraint Equ.~(\ref{eq:constraintMass}) and commute with the ADM Hamiltonian are easily constructed. An explicit example is the following. Given a scalar matter field operator $\hat{O}(x)$, we can construct the physical operator
\begin{equation}
    \hat{O}^{\text{Phys}}(x)=\hat{O}(x+\sqrt{16\pi G_{N}}V(x))\,,\label{eq:Ophys}
\end{equation}
for which we have replaced the coordinate $x^{\mu}$ in $\hat{O}(x)$ by the invariant combination $x^{\mu}+\sqrt{16\pi G_{N}}V^{\mu}(x)$. Using Equ.~(\ref{eq:canonicalV0}), it is straightforward to check that we have
\begin{equation}
[\sqrt{16\pi G_{N}}\hat{H}_{\text{QFT}}-\hat{\Pi}_{V^{0}},\hat{\mathcal{O}}^{\text{Phys}}(x)]=0\,,\quad\text{and }[\hat{H}_{\partial},\hat{\mathcal{O}}^{\text{Phys}}(x)]=0\,.
\end{equation}
In fact, the operator Equ.~(\ref{eq:Ophys}) is invariant under all diffeomorphisms to the leading nontrivial order in $G_{N}$. This can be seen by noticing that at the leading order in $G_{N}$, the diffeomorphism transform of the Goldstone vector field Equ.~(\ref{eq:diffV}) is
\begin{equation}
    V^{\mu}(x)\rightarrow V^{\mu}(x)-\epsilon^{\mu}(x)\,,\quad\text{as }x^{\mu}\rightarrow x^{\mu}+\sqrt{16\pi G_{N}}\epsilon^{\mu}(x)\,.
\end{equation}
Therefore, the combination $x^{\mu}+\sqrt{16\pi G_{N}}V^{\mu}(x)$ is invariant under the diffeomorphism transform. Similar constructions can be carried out for operators with nonzero spins. Last but not least, we notice that the momentum constraints are modified to
\begin{equation}
    \mathcal{H}_{i}=-2g_{ij}D_{k}\Pi^{jk}+\mathcal{H}_{i,\text{matter}}-\frac{1}{\sqrt{16 \pi G_{N}}}\pi_{V^{i}}=0\,,
\end{equation}
where we used the fact that the canonical momentum of the $i$-th component of $V^{\mu}(x)$ is
\begin{equation}
   \pi_{V^{i}}(x)= M^{2}\tilde{h}_{0}^{i}(x)\,.
\end{equation}
Thus, the operator Equ.~(\ref{eq:Ophys}) also commutes with the modified momentum constraint. In summary, the basic consistency of entanglement wedge reconstruction identified in \cite{Geng:2021hlu} requires that operators in the entanglement island should be of the form Equ.~(\ref{eq:Ophys}). The fact that island model is in the Higgs phase with a massive graviton enables this consistency condition to be obeyed. We should also notice that similar to \cite{Chandrasekaran:2022cip,Witten:2023qsv} the Hamiltonian constraint is modified by an additional term and this additional term allows localized operators in a closed gravitational region. However, in our case—unlike approaches that introduce an observer as an external input to the theory—this additional term arises dynamically from the bath coupling which spontaneously breaks the diffeomorphism symmetry.

More interestingly, it is recently pointed out in \cite{Geng:2025rov} that there is another basic consistency condition for operators in the entanglement island to be with the holographic interpretation of entanglement islands. This is based on the observation that the operators Equ.~(\ref{eq:Ophys}) in the island are localized in time so they necessarily created energetic excitations. Since they are fully encoded by the bath subregion $R$, they should at least have a nonzero commutator with the bath Hamiltonian $\hat{H}_{\text{bath}}$. This consistency condition is obeyed due to the fact that the Goldstone boson $V^{\mu}(x)$ is a composite operator, which can be seen from Equ.~(\ref{eq:VUdual}) and Equ.~(\ref{eq:Udef}), which consists of both the bath operator and the dual CFT$_{d}$ operator. Thus, $V^{\mu}(x)$ itself doesn't commute with the bath Hamiltonian and this induces a nonzero commutator between the island operators Equ.~(\ref{eq:Ophys}) and $\hat{H}_{\text{bath}}$. For more details on this nonzero commutator, we refer the readers to \cite{Geng:2025rov}.

\section{Operator Algebras in Entanglement Islands}\label{sec:algebra}
From the previous section, we see that operators inside the island should be of the form Equ.~(\ref{eq:Ophys}). Let's denote the set of such operators to be $\mathcal{A}_{I}$. As we have seen that operators in $\mathcal{A}_{I}$ obey the diffeomorphism constraints, they define a gravitational algebra. Nevertheless, this gravitational algebra is so far not different from the algebra of quantum field theory operators. Hence $\mathcal{A}_{I}$ is a Type III$_{1}$ von Neumann algebra which doesn't have a trace. In this section, we will first show that with an appropriate element added to this algebra, it becomes a Type II$_{\infty}$ von Neumann factor which should have a trace. Then we construct the trace of this algebra and study the associated entropy. We argue that the resulting entropy is the generalized gravitational entropy of the entanglement island and it should be UV finite.

\subsection{Operator Algebras in Islands as Crossed Products}\label{sec:crossprod}

We note that in the algebra $\mathcal{A}_{I}$ we didn't add the operators constructed purely out of the Goldstone boson $V^{\mu}(x)$ and its canonical conjugate $\pi_{V^{\mu}}(x)$. Unlike $V^{\mu}(x)$, the operators $\pi_{V^{\mu}}(x)$ obeys all diffeomorphism constraints. In fact, we can conjoin the algebra $\mathcal{A}_{I}$ with any operator $\Pi[\xi]$ of the form
\begin{equation}
    \Pi[\xi]=\frac{1}{\sqrt{16\pi G_{N}}}\int_{\Sigma} d^{d}x\sqrt{-g^{0}}N \xi^{\mu}(x)\pi_{V^{\mu}}(x)\,,\label{eq:Pixi}
\end{equation}
where $\xi^{\mu}(x)$ is a vector field and the integral is taken to be along the Cauchy slice $\Sigma$ of the AdS$_{d+1}$ on which the island $I$ resides. It is well-motivated to consider vector fields $\xi^{\mu}(x)$ that do not deform the entanglement island, i.e. they leave the domain of dependence $D(I)$ invariant. With any such operator added, the algebra $\mathcal{A}_{I}$ is deformed into a crossed product $\mathcal{A}_{I}\rtimes\mathcal{A}_{\Pi[\xi]}$, where $\mathcal{A}_{\Pi[\xi]}$ is the algebra generated by a single element $\hat{\Pi}[\xi]$. This algebra is in fact unitarily equivalent to $\mathcal{A}_{I,\text{QFT}}\rtimes\mathcal{A}_{\Pi[\xi]+H_{\text{QFT}}[\xi]}$, 
where $\mathcal{A}_{I,\text{QFT}}$ denotes the algebra of quantum field theory operators inside the entanglement island $I$.\footnote{We note that $A_{I,\text{QFT}}$ is in fact the same as $A_{D(I),\text{QFT}}$ with $D(I)$ the domain of dependence of the spatial subregion $I$ \cite{Witten:2023qsv}.} This can be seen by implementing the following unitarity transform on to the original algebra $\mathcal{A}_{I}\rtimes\mathcal{A}_{\Pi[\xi]}$
\begin{equation}
    \hat{U}[V]=e^{i\sqrt{16\pi G_{N}}\hat{H}_{\text{QFT}}[V]}\,,
\end{equation}
for which we have
\begin{equation}
     \hat{H}_{\text{QFT}}[V]=\int_{\Sigma} d^{d}x\sqrt{-g^{0}} N V^{\mu}(x) \mathcal{H}_{\text{QFT},\mu}(x)\,,
\end{equation}
where $\mathcal{H}_{\text{QFT},\mu}(x)$ denotes the matter stress-energy density as appeared in the constraints Equ.~(\ref{eq:constraints}), with the integral along Cauchy slice that the island $I$ resides. More precisely, we have
\begin{equation}
    \mathcal{A}_{I,\text{QFT}}\rtimes\mathcal{A}_{\Pi[\xi]+H_{\text{QFT}}[\xi]}=\hat{U}^{\dagger}[V]\Big(\mathcal{A}_{I}\rtimes\mathcal{A}_{\Pi[\xi]}\Big)\hat{U}[V]\,,
\end{equation}
where we note that the effect of the unitary transform $\hat{U}[V]$ on $\mathcal{A}_{I}$ is to undress the operators and so we have $\mathcal{A}_{I,\text{QFT}}$ on the left hand side.

We will assume the so-called geometric modular flow conjecture \cite{Jensen:2023yxy}, which states that for the Hilbert space $\mathcal{H}_{\text{QFT}}$ that the algebra $\mathcal{A}_{I,\text{QFT}}$ acts on, there is a cyclic and separating state $\ket{\Psi}$ whose modular flow is geometric and leaves $\partial I$ invariant. That is there exists a vector field $\xi^{\mu}_{\Psi}(x)$ which doesn't deform $D(I)$ and it satisfies
\begin{equation}
    \Delta_{\Psi}=e^{-H_{\text{QFT}}[\xi_{\Psi}]}\,,
\end{equation}
along the Cauchy slice $\Sigma$. In the above formula $\Delta_{\Psi}$ is the modular operator associated with the state $\ket{\Psi}$, i.e. $H_{\text{QFT}}[\xi_{\Psi}]$ is the modular Hamiltonian of the state $\ket{\Psi}$. Meanwhile, this vector field approaches to the global time translation at the asymptotic boundary \cite{Jensen:2023yxy}. We will conjoin the algebra with $\Pi[\xi_{\Psi}]$ for any such $\ket{\Psi}$. The above analysis shows that this new algebra is unitarily equivalent to
\begin{equation}
    \mathcal{A}=\mathcal{A}_{I,\text{QFT}}\rtimes\mathcal{A}_{\Pi[\xi_{\Psi}]+H_{\text{QFT}}[\xi_{\Psi}]}\,,
\end{equation}
which is exactly the crossed product of the QFT algebra $\mathcal{A}_{I,\text{QFT}}$ by its modular automorphism group. Thus, this new algebra $\mathcal{A}$ is a Type II$_{\infty}$ von Neumann factor \cite{10.1007/BF02392041,Witten:2021unn}. More precisely, the Hilbert space that $\mathcal{A}$ acts on is $\mathcal{H}_{\text{QFT}}\otimes L^{2}(\mathbb{R})$, where $L^{2}(\mathbb{R})$ are the set of square integrable functions of $\mathbb{R}$.\footnote{$\mathbb{R}$ is isomorphic to the spectrum of the operator $\hat{\Pi}[\xi_{\Psi}]$.} A general element of the new algebra $\mathcal{A}$ is an $\mathcal{A}_{I,\text{QFT}}$-valued function of $\hat{\Pi}[\xi_{\Psi}]+\hat{H}_{\text{QFT}}[\xi_{\Psi}]$.

\subsection{Trace and Density Matrices of the Algebra}\label{sec:trace}
In this subsection, we will work out the trace and the density matrix of the algebra $\mathcal{A}$ we constructed in Sec.~\ref{sec:crossprod}. We will follow the constructions in \cite{Witten:2021unn,Chandrasekaran:2022cip,Chandrasekaran:2022eqq,Jensen:2023yxy}.

Let's first work out a few details about the algebra $\mathcal{A}$. This algebra can be written in the following form
\begin{equation}
    \mathcal{A}=\left\{ a, e^{i(Y+\hat{H}_{\Psi})t}\middle| \forall a\in \mathcal{A}_{I,\text{QFT}},t\in \mathbb{R} \right\}''\simeq\mathcal{A}_{I,\text{QFT}}\times\mathcal{B}(L^{2}(\mathbb{R}))\,,\label{eq:defA}
\end{equation}
where for convenience we have defined $Y=\Pi[\xi_{\Psi}]=-\hat{H}_{\text{obs}}$ and $\hat{H}_{\Psi}=\hat{H}_{\text{QFT}}[\xi_{\Psi}]$, and $\mathcal{B}(L^{2}(\mathbb{R}))$ denotes the algebra of all bounded operators on the space $L^{2}(\mathbb{R})$. The $''$ in the above equation denotes double commutant, which produces a closed set under the operator norm topology. A generic element of $\mathcal{A}$ is of the form $ae^{i(Y+\hat{H}_{\Psi})t}$, where $a\in \mathcal{A}_{I,\text{QFT}}$. To check that $\mathcal{A}$ is indeed an algebra, let's consider the product of any two elements of $\mathcal{A}$
\begin{equation}
    \begin{split}
        a e^{i(Y+\hat{H}_{\Psi})t}be^{i(Y+\hat{H}_{\Psi})s}&=a e^{i(Y+\hat{H}_{\Psi})t}be^{-i(Y+\hat{H}_{\Psi})t}e^{i(Y+\hat{H}_{\Psi})(t+s)}\\&=a e^{i\hat{H}_{\Psi}t}be^{-i\hat{H}_{\Psi}t}e^{i(Y+\hat{H}_{\Psi})(t+s)}\\&=ab'e^{i(Y+\hat{H}_{\Psi})(t+s)}\in\mathcal{A}\,,
    \end{split}
\end{equation}
where we have used the fact that $\hat{H}_{\Psi}$ is the matter modular Hamiltonian which generates an outer automorphism of $\mathcal{A}_{I,\text{QFT}}$. Thus, $\mathcal{A}$ is indeed an algebra.

Now let's consider the trace for the algebra $\mathcal{A}$. The state $\ket{\Psi}$ we are considering belongs to the representation space $\mathscr{H}_{\text{QFT}}$ of $\mathcal{A}_{I,\text{QFT}}$. So to construct the trace for $\mathcal{A}$, let's first take a state in $\mathscr{H}_{\text{QFT}}\otimes L^{2}(\mathbb{R})$ which is the representation space of $\mathcal{A}$
\begin{equation}
    |\widehat{\Psi}\rangle=\ket{\Psi}\otimes g^{\frac{1}{2}}(Y)\,,
\end{equation}
where $g(Y)$ is everywhere positive which, together with $|\Psi\rangle$ being cyclic and separating, ensures that $|\widehat{\Psi}\rangle$ is cyclic and separating \cite{Witten:2021unn}. Following \cite{Witten:2021unn}, we have the modular operator associated with the state $|\widehat{\Psi}\rangle$ as
\begin{equation}
    \widehat{\Delta}_{\widehat{\Psi}}=\Delta_{\Psi}\cdot g(Y+\hat{H}_{\Psi}) (g(Y))^{-1}\,, 
\end{equation}
which can be factorized as
\begin{equation}
    \widehat{\Delta}_{\widehat{\Psi}}=\tilde{K}K\,,\label{eq:factorization}
\end{equation}
with
\begin{equation}
    K=\Delta_{\Psi}e^{-Y}g(Y+\hat{H}_{\Psi})=e^{-(Y+\hat{H}_{\Psi})}g(Y+\hat{H}_{\Psi})\,,\quad \tilde{K}=\frac{e^{Y}}{g(Y)}\,.
\end{equation}
If we take $g(Y)$ such that $e^{-Y}g(Y)\in\mathcal{B}(L^{2}(\mathbb{R}))$, then the factorization Equ.~(\ref{eq:factorization}) has the nice property that
\begin{equation}
    K\in \mathcal{A}\,,\quad\tilde{K}\in \mathcal{A}'\,,
\end{equation}
where $\mathcal{A}'$ denotes the commutant of $\mathcal{A}$. This property enables us to define the trace of $\mathcal{A}$ as 
\begin{equation}
    \Tr \hat{a}\equiv\langle\widehat{\Psi}|\hat{a}K^{-1}|\widehat{\Psi}\rangle\,,
\end{equation}
for any $\hat{a}\in\mathcal{A}$. It is straightforward to check that the above trace is indeed tracial.

\begin{equation}
\begin{split}
    \Tr \hat{a}\hat{b}&=\langle\widehat{\Psi}|\hat{a}\hat{b}K^{-1}|\widehat{\Psi}\rangle=\langle\widehat{\Psi}|\hat{b}K^{-1}\widehat{\Delta}_{\widehat{\Psi}}\hat{a}|\widehat{\Psi}\rangle\\&=\langle\widehat{\Psi}|\hat{b}K^{-1}\widehat{\Delta}_{\widehat{\Psi}}\hat{a}\widehat{\Delta}^{-1}_{\widehat{\Psi}}|\widehat{\Psi}\rangle\\&=\langle\widehat{\Psi}|\hat{b}K^{-1}K\tilde{K}\hat{a}\tilde{K}^{-1}K^{-1}|\widehat{\Psi}\rangle\\&=\langle\widehat{\Psi}|\hat{b}\hat{a}K^{-1}|\widehat{\Psi}\rangle=\Tr \hat{b}\hat{a}\,,
    \end{split}
\end{equation}
where we used the facts that $\forall\hat{a},\hat{c}\in \mathcal{A}$ we have $\langle\widehat{\Psi}|\hat{a}\hat{c}|\widehat{\Psi}\rangle=\langle\widehat{\Psi}|\hat{c}\widehat{\Delta}_{\widehat{\Psi}}\hat{a}|\widehat{\Psi}\rangle$ and $|\widehat{\Psi}\rangle$ is invariant under its own modular follow generated by $\widehat{\Delta}_{\widehat{\Psi}}$. This last fact is essentially the KMS property for thermal states. Also, it is easy to see that the trace $\Tr \hat{a}$ for any $\hat{a}\in\mathcal{A}$ is in fact independent of $g(Y)$ \cite{Witten:2021unn}.\footnote{This can be done by noticing that $\hat{H}_{\Psi}\ket{\Psi}=0$.}

Having constructed the trace, we are ready to write down the density matrix of the algebra $\mathcal{A}$ for any semiclassical state $|\widehat{\Phi}\rangle\in \mathscr{H}_{\text{QFT}}\otimes L^{2}(\mathbb{R})$ where there could exist faithful islands. For semiclassical quantum states, the bulk has a clear geometrical description. A necessary condition for the island to be able to faithfully exist is that operators could be faithfully localized inside the island. We have identified the existence of the Goldstone vector $V^{\mu}(x)$ as a necessarily condition for operators to be localized inside the island. Thus, the Goldstone vector field $V^{\mu}(x)$ must act as a faithful operator with small variations in order for operators to be faithfully localized within the island. In other words, the conjugate variable $Y=\Pi[\xi_{\Psi}]$ should have a large fluctuation in the semiclassical states where island could faithfully exist. We will call this condition the \textit{faithful island condition}.\footnote{An interesting question is to understand if all semiclassical states obey the faithful island condition.} However, we should notice that the faithful island condition should be imposed on the states which are representations of the original algebra $\mathcal{A}_{I}\rtimes\mathcal{A}_{\Pi[\xi]}=\hat{U}[V]\mathcal{A}\hat{U}^{\dagger}[V]$. This original algebra obeys the diffeomorphism constraint, and so are the representations of this original algebra (see also  \cite{Jensen:2023yxy} for relevant discussions). Thus, for the algebra $\mathcal{A}$ we will consider the states 
\begin{equation}
    |\widehat{\Phi}\rangle=\hat{U}^{\dagger}[V]\ket{\Phi}\ket{f}=\int_{-\infty}^{\infty} dy\sqrt{\epsilon}f(\epsilon y)\hat{U}^{\dagger}[V]\ket{\Phi}\ket{y}\,,\label{eq:Phi}
\end{equation}
for arbitrary $\ket{\Phi}\in \mathscr{H}_{\text{QFT}}$ and normalized $f(x)\in L^{2}(\mathbb{R})$ with $\epsilon\ll1$. For such states, we have the density matrix
\begin{equation}
    \rho_{\widehat{\Phi}}=\epsilon \bar{f}\big(\epsilon (Y+\hat{H}_{\Psi})\big)e^{-Y}\Delta_{\Phi|\Psi}f\big(\epsilon(Y+\hat{H}_{\Psi})\big)\,,\label{eq:densitymatrix}
\end{equation}
where $\Delta_{\Phi|\Psi}$ is the relative modular operator for the two states $\ket{\Phi},\ket{\Psi}\in \mathscr{H}_{\text{QFT}}$. Following \cite{Chandrasekaran:2022eqq}, it is easy to check that we have
\begin{equation}
    \Tr \rho_{\widehat{\Phi}}\hat{a}=\langle\widehat{\Phi}|\hat{a}|\widehat{\Phi}\rangle\,,\quad\forall\hat{a}\in\mathcal{A}\,,
\end{equation}
and $\rho_{\widehat{\Phi}}\in\mathcal{A}$.\footnote{For the last purpose, one can notice that $e^{-Y}=e^{-(Y+\hat{H}_{\Psi})}e^{\hat{H}_{\Psi}}=e^{-(Y+\hat{H}_{\Psi})}\Delta^{-1}_{\Psi}$. Thus, we have
\begin{equation}
    \rho_{\widehat{\Phi}}=\epsilon \bar{f}\big(\epsilon (Y+\hat{H}_{\Psi})\big)\Delta_{\Phi|\Psi}\Delta^{-1}_{\Psi}e^{-(Y+\hat{H}_{\Psi})}f\big(\epsilon(Y+\hat{H}_{\Psi})\big)\,.
\end{equation}
Using the fact that Connes cocyle flow $u_{\Phi|\Psi}=\Delta_{\Phi|\Psi}^{is}\Delta^{-is}_{\Psi}\in\mathcal{A}_{I,QFT}$ ((see Appendix.~\ref{sec:review1})), we can imply that $\rho_{\widehat{\Phi}}\in\mathcal{A}$. } We emphasize that the choice $\epsilon\ll1$ ensures that the state $\hat{U}^{\dagger}[V]|\widehat{\Phi}\rangle$ satisfies the faithful island condition \cite{Chandrasekaran:2022eqq}, and meanwhile this state obeys all diffeomorphism constraints.

\subsection{Calculating the Entropy of the Algebra}
With the trace and the density matrix equipped, we can compute the entropy of the algebra $\mathcal{A}$ as we defined in Equ.~(\ref{eq:defA}). The entropy is
\begin{equation}
    S(\rho_{\widehat{\Phi}})=-\Tr \rho_{\widehat{\Phi}}\log\rho_{\widehat{\Phi}}=-\langle\widehat{\Phi}|\log\rho_{\widehat{\Phi}}|\widehat{\Phi}\rangle\,.\label{eq:entropy1}
\end{equation}
We note that in general the modular Hamiltonian $\hat{H}_{\Psi}$ and the relative modular operator $\Delta_{\Phi|\Psi}$ don't commute. However, the function $f(\epsilon(Y+\hat{H}_{\Psi}))$ is slowly changing due to $\epsilon\ll1$. Thus, we have
\begin{equation}
    [f(\epsilon(Y+\hat{H}_{\Psi})),\Delta_{\Phi|\Psi}]=\mathcal{O}(\epsilon)\,.
\end{equation}
Therefore, given Equ.~(\ref{eq:densitymatrix}), we have
\begin{equation}
    -\log\rho_{\widehat{\Phi}}=Y-\log\Delta_{\Phi|\Psi}-\log\big(\epsilon\abs{f(\epsilon(Y+\hat{H}_{\Psi}))}^2\big)\,,
\end{equation}
where we ignored $\mathcal{O}(\epsilon)$ terms as we only consider states that obey the faithful island condition. As a result, the entropy Equ.~(\ref{eq:entropy1}) can be written as
\begin{equation}
    S(\rho_{\widehat{\Phi}})=\langle Y\rangle_{f}-\bra{\Phi}\hat{H}_{\Psi}\ket{\Phi}-\langle\widehat{\Phi}|\log\Delta_{\Phi|\Psi}|\widehat{\Phi}\rangle-\langle\widehat{\Phi}|\log\big(\epsilon\abs{f(\epsilon (Y+\hat{H}_{\Psi}))}^2\big)|\widehat{\Phi}\rangle\,,\label{eq:entropy2}
\end{equation}
where again we ignored $\mathcal{O}(\epsilon)$ terms. The third term in Equ.~(\ref{eq:entropy2}) can be calculated if one uses $\Delta_{\Phi|\Psi}=\rho_{\Phi}\otimes\rho'^{-1}_{\Psi}$ where $\rho'_{\Psi}$ is the QFT density matrix for the complementary region of the island inside the gravitational AdS for the state $\ket{\Psi}$\footnote{We note that strictly speaking $\Delta_{\Phi|\Psi}$ is the modular operator for a Type III$_{1}$ algebra which doesn't allow a factorization as written. This is because neither $\rho_{\Phi}$ nor $\rho'_{\Psi}$ is a strictly speaking well-defined quantity for QFT due to the strong localized entanglement at $\partial I$. Nevertheless, this decomposition is safe as $\Delta_{\Phi|\Psi}$ is well-defined and this splitting is only helpful to rewrite the result in a physically interpretable form. Thus, we will do this sloppy decomposition hereafter, and keep in mind that $S(\rho_{\widehat{\Phi}})$ is well-defined and finite, no matter how we decompose it.}. Thus, we have
\begin{equation}
\begin{split}
    -\langle\widehat{\Phi}|\log\Delta_{\Phi|\Psi}|\widehat{\Phi}\rangle&= -\bra{f}\langle\Phi|\hat{U}[V]\log\Delta_{\Phi|\Psi}\hat{U}^{\dagger}[V]|\Phi\rangle\ket{f}\\&=-\bra{f}\langle\Phi|\log\Delta_{\Phi|\Psi}|\Phi\rangle\ket{f}+\mathcal{O}(\epsilon)\\&=S(I)_{\text{QFT},\Phi}+\langle\Phi|\int_{\bar{I}} d^{d}x\sqrt{-g^{0}} N \xi_{\Psi}^{\mu}(x) \mathcal{H}_{\text{QFT},\mu}(x)|\Phi\rangle
    \end{split}
\end{equation}
The last term in Equ.~(\ref{eq:entropy2}) can be simplified by noticing that
\begin{equation}
    \begin{split}
 \log\big(\epsilon\abs{f(\epsilon (Y+\hat{H}_{\Psi}))}^2\big)&=\log\big(\epsilon \hat{U}^{\dagger}[V]\abs{f(\epsilon Y)}^2 \hat{U}[V]\Big)\\&= \hat{U}^{\dagger}[V]\log\big(\epsilon\abs{f(\epsilon Y)}^2\big)\hat{U}[V]\,,
    \end{split}
\end{equation}
where we used the fact that for a matrix the unitary transform can be pulled outside of the logarithm. Thus, we have
\begin{equation}
    \begin{split}
        \langle\widehat{\Phi}|\log\big(\epsilon\abs{f(\epsilon (Y+\hat{H}_{\Psi}))}^2\big)|\widehat{\Phi}\rangle&=\langle\log\big(\epsilon\abs{f(\epsilon Y)}^2\big)\rangle_{f}\,.
    \end{split}
\end{equation}
In summary, the entropy can be written in the following form
\begin{equation}
\begin{split}
S(\rho_{\widehat{\Phi}})=\langle Y\rangle_{f}+S(I)_{\text{QFT},\Phi}-\langle\Phi|\int_{I} d^{d}x\sqrt{-g^{0}} N \xi_{\Psi}^{\mu}(x) \mathcal{H}_{\text{QFT},\mu}(x)|\Phi\rangle-\langle\log\big(\epsilon\abs{f(\epsilon Y)}^2\big)\rangle_{f}\,,\label{eq:entropy3}
\end{split}
\end{equation}
where we used $\bra{\Phi}\hat{H}_{\Psi}\ket{\Phi}=\langle\Phi|\int_{\Sigma} d^{d}x\sqrt{-g^{0}} N \xi_{\Psi}^{\mu}(x) \mathcal{H}_{\text{QFT},\mu}(x)|\Phi\rangle$.

\subsection{The Entropy as the Generalized Entropy}
As we discussed, the state $\ket{\Phi}\ket{f}$ obeys all constraints. We can use this fact to compute the third term in the entropy Equ.~(\ref{eq:entropy3}). We will do this simplification by a direct calculation in our case and the result can be easily covariantized.

In our case, we have an empty AdS in the Poincar\'{e} patch as the background geometry. We are considering the Cauchy slice $\Sigma$ to be a constant-time slice. The vector field $\xi_{\Psi}^{\mu}(x)$ generates the modular flow of the state $\ket{\Psi}$ on this Cauchy slice $\Sigma$ with $\partial I$ fixed, so we have
\begin{equation}
    \xi^{\mu}_{\Psi}(x)|_{\partial I}=0\,,\quad \nabla_{\mu}\xi_{\nu\Psi}(x)|_{\partial I}=2\pi n_{\mu\nu}\,,
\end{equation}
where $\nabla_{\mu}$ is the covariant derivative with respect to the background spacetime and $n_{\mu\nu}$ is the unit binormal to $\partial I$ in the background spacetime. Moreover, $\xi_{\Psi}^{\mu}(x)$ is future directed inside the island $I$, past directed outside the island, and coincides with the global (pass directed) time translation near the asymptotic boundary $\partial\Sigma$. Since we are studying gravitational perturbation theory, we can simplify the third term in Equ.~(\ref{eq:entropy3}) by linearizing the constraints Equ.~(\ref{eq:constraints}). The Hamiltonian constraint has already been linearized in Equ.~(\ref{eq:core}) and Equ.~(\ref{eq:HADM}). Thus, let's first linearize the momentum constraint. To the first nontrivial order and before including the contribution from the Goldstone vector field, we have
\begin{equation}
\mathcal{H}_{i}^{(1)}=\frac{1}{\sqrt{16\pi G_{N}}N}\Big(g^{(0)nk}\partial_{k}\dot{h}_{in}-g_{ij}^{(0)}\partial_{k}(g^{(0)jk}g^{(0)mn}\dot{h}_{mn})-zg^{(0)}_{iz}g^{(0)mn}\dot{h}_{mn}-(d+2)z\dot{h}_{iz}\Big)+\mathcal{H}_{i,QFT}\,.
\end{equation}
Including the contributions from the Goldstone vector field and imposing the constraint to leading order, we have
\begin{equation}
    \begin{split}
&\int_{I} d^{d}x\sqrt{-g^{0}} N \xi_{\Psi}^{\mu}(x) \mathcal{H}_{\text{QFT},\mu}(x)\\&=\int_{\partial I} d^{d-1}x\sqrt{g^{0}_{d-2}}2\pi\frac{h^{a}_{a}}{\sqrt{16\pi G_{N}}}+\frac{1}{\sqrt{16\pi G_{N}}}\int_{I} d^{d}x\sqrt{-g^{0}}N \xi_{\Psi}^{\mu}(x)\pi_{V^{\mu}}(x)+\frac{1}{\sqrt{16\pi G_{N}}}F_{\Psi}[h,\dot{h}]\,,
    \end{split}
\end{equation}
where $g^{0}_{d-2}$ is the determinant of the induced metric on $\partial I$ in the background spacetime, $h^{a}_{a}$ is the trace of the fluctuation of the induced metric on $\partial I$, and $F_{\Psi}[h,\dot{h}]$ is a single integral over $\bar{I}$ with the integrand linear in the linearized metric fluctuations and independent of their spatial derivatives. In fact, we have
\begin{equation}
    \delta A(\partial I)=\delta\int_{\partial I} d^{d-1}x\sqrt{g_{d-2}}=\frac{\sqrt{16\pi G_{N}}}{2}\int_{\partial I} d^{d-1}x\sqrt{g^{0}_{d-2}}h^{a}_{a}\,, 
\end{equation}
which implies that
\begin{equation}
\begin{split}
   & \int_{I} d^{d}x\sqrt{-g^{0}} N \xi_{\Psi}^{\mu}(x) \mathcal{H}_{\text{QFT},\mu}(x)\\&=\frac{\delta A(\partial I)}{4 G_{N}}+\frac{1}{\sqrt{16\pi G_{N}}}\int_{I} d^{d}x\sqrt{-g^{0}}N \xi_{\Psi}^{\mu}(x)\pi_{V^{\mu}}(x)+\frac{1}{\sqrt{16\pi G_{N}}}F_{\Psi}[h,\dot{h}]\,.\label{eq:barIconstraint}
    \end{split}
\end{equation} 
We also follow \cite{Jensen:2023yxy} making the following gauge choice
\begin{equation}
    F_{\Psi}[h,\dot{h}]=0\,,\label{eq:gc}
\end{equation}
As a result, we have 
\begin{equation}
\langle\widehat{\Phi}|\int_{I} d^{d}x\sqrt{-g^{0}} N \xi_{\Psi}^{\mu}(x) \mathcal{H}_{\text{QFT},\mu}(x)|\widehat{\Phi}\rangle=\frac{\delta A(\partial I)}{4 G_{N}}+\langle\widehat{\Phi}|\int_{I} d^{d}x\sqrt{-g^{0}}N \xi_{\Psi}^{\mu}(x)\pi_{V^{\mu}}(x)|\widehat{\Phi}\rangle
    \,.\label{eq:replace}
\end{equation} 
Substituting Equ.~(\ref{eq:replace}) into Equ.~(\ref{eq:entropy3}), we have
\begin{equation}
\begin{split}
S(\rho_{\widehat{\Phi}})=S(I)_{\text{QFT},\Phi}+\langle\frac{\delta A(\partial I)}{4 G_{N}}\rangle_{\Phi}+\langle\widehat{\Phi}|\int_{\bar{I}} d^{d}x\sqrt{-g^{0}}N \xi_{\Psi}^{\mu}(x)\pi_{V^{\mu}}(x)|\widehat{\Phi}\rangle-\langle\log\big(\epsilon\abs{f(\epsilon Y)}^2\big)\rangle_{f}
\,,\label{eq:entropy4}
\end{split}
\end{equation}
for which we can identify the last two terms as the contribution of the entropy from the ``observer"
\begin{equation}
    S_{\text{obs},f}=\langle\widehat{\Phi}|\int_{\bar{I}} d^{d}x\sqrt{-g^{0}}N \xi_{\Psi}^{\mu}(x)\pi_{V^{\mu}}(x)|\widehat{\Phi}\rangle-\int_{-\infty}^{\infty} dy \epsilon\abs{f(\epsilon y)}^2\log\big(\epsilon\abs{f(\epsilon y)}^2\big)\,.
\end{equation}
Thus, up to a state-independent constant $c$ we have
\begin{equation}
    S(\rho_{\widehat{\Phi}})=\langle\frac{ A(\partial I)}{4 G_{N}}\rangle_{\Phi}+S(I)_{\text{QFT},\Phi}+S_{\text{obs},f}-c\,,
\label{eq:entropy5}
\end{equation}
where $c$ can be thought of as the area of $\partial I$ in Planck units in the background geometry.

In summary, under the gauge choice Equ.~(\ref{eq:gc}) made in \cite{Jensen:2023yxy}, the entropy we derived for the algebra $\mathcal{A}$ is in fact the generalized entropy of the island, together with the observer's contribution, up to a state-independent constant. We should notice that in the above formula $G_{N}$ is not the bare Newton's constant $G_{N,0}$ but the Newton's constant renormalized by the transparent matter fields, which we have integrated out at the very beginning. Furthermore, the entropy $S(I)_{\text{QFT},\Phi}$ is the entanglement entropy of the reflecting matter fields inside the island. The entropy $S(\rho_{\widehat{\Phi}})$ should be finite as it is the entropy of a Type II$_{\infty}$ von Neumann algebra. We should emphasize that to get the form of the generalized entropy it is crucial to make the gauge choice Equ.~(\ref{eq:gc}). This gauge choice is automatically satisfied if $\xi_{\Psi}^{\mu}(x)$ is a Killing vector field. We leave a better understanding of the gauge choice Equ.~(\ref{eq:gc}) for general $\xi_{\Psi}^{\mu}(x)$ and the ``observer" contribution $S_{\text{obs},f}$ to future work.

\section{Comments on Type II$_{1}$ Constructions}\label{sec:comment}
We note that the trace we defined in Sec.~\ref{sec:trace} is only well-defined for a class of operators in $\mathcal{A}$. This class is called \textit{trace class}. Operators outside the trace class include the identity operator. This is because
\begin{equation}
    \Tr \mathbbm{1}=\langle\widehat{\Psi}|K^{-1}|\widehat{\Psi}\rangle=\int_{-\infty}^{\infty} dy e^{-Y}=\infty\,.
\end{equation}
Thus, the trace cannot be conventionally normalized and there always exists an ambiguity in this trace as an overall factor. This ambiguity also manifests in the factorization Equ.~(\ref{eq:factorization}) where
\begin{equation}
    K\rightarrow e^{-c}K\,,\quad\tilde{K}\rightarrow e^{c}\tilde{K}\,,
\end{equation}
is also a valid factorization and under this rescaling, the trace transforms as
\begin{equation}
    \Tr\rightarrow e^{c}\Tr\,.
\end{equation}
This ambiguity explains the fact that the entropy is only defined up to a state-independent constant.

To ameliorate the situation against the above observations and push the algebra closer to the Type I von Neumann algebra, the authors of \cite{Jensen:2023yxy} following \cite{Chandrasekaran:2022cip,Witten:2023qsv} proposed that one can consider projecting the ``observer" Hamiltonian to be bounded from below. With such a projection $P_{0}$ incorporated, the trace of the identity becomes
\begin{equation}
    \Tr \mathbbm{1}=\langle\widehat{\Psi}|P_{0} K^{-1}P_{0}|\widehat{\Psi}\rangle=\int_{-\infty}^{0} dy e^{Y}=1\,,
\end{equation}
where we remember that $\hat{H}_{\text{obs}}=-Y$. Thus, the trace is conventionally normalized and the ambiguity in the entropy can be removed. In fact, with this projection, the algebra $\mathcal{A}$ becomes Type II$_{1}$ which admits a normalized trace and a finite entropy with no ambiguity. The result is especially tantalizing in the context of de Sitter holography \cite{Karch:2003em,Alishahiha:2004md,Dong:2018cuv,Geng:2019bnn,Geng:2019ruz,Geng:2020kxh,Geng:2021wcq}. As it was noticed in \cite{Chandrasekaran:2022cip}, the Type II$_{1}$ von Neumann might be helpful to explain the observation in \cite{Geng:2019ruz,Geng:2020kxh} that the density matrix of the de Sitter static patch is a maximally entangled state proportional to the identity operator
\begin{equation}
    \rho_{\text{dS static}}=e^{-S_{\text{GH}}}\mathbbm{1}\,,
\end{equation}
where $S_{\text{GH}}$ is the de Sitter Gibbons-Hawking entropy \cite{Gibbons:1976ue}. This is because the Type II$_{1}$ von Neumann algebra admits a maximally entangled state of exactly the same form due to the normalizability of the identity operator under the trace.

The physical argument for this projection is that the ``observer" Hamiltonian $\hat{H}_{\text{obs}}$ should be bounded from below as for any quantum system. However, as we have discussed, as opposed to standard quantum systems, the observer Hamiltonian is linear in a phase space variable. Hence, one shouldn't expect the general properties of standard quantum systems to be straightforwardly carried over to the ``observer". This is especially the case in our explicit model for the ``observer". In our model, the ``observer" is a particular mode of the Goldstone boson which is not bounded from below. The emergence of the observer is due to the spontaneous breaking of the AdS diffeomorphisms by the bath. This spontaneous symmetry breaking modifies the diffeomorphism constraint and the modification is the addition of these ``observer" term. Such a modification enables diffeomorphism invariant operators to be localized inside the island. In fact, as pointed out in \cite{Geng:2025rov}, the emergent ``observer" term plays a crucial role for the internal consistency of the holographic interpretation of entanglement islands. In a sense, the observer is a mode transferring energy between the gravitational AdS and the bath. Energy can flow either way, so there is no reason for the ``observer" Hamiltonian to be bounded from below. Hence, we believe the projection of the ``observer" Hamiltonian postulated in \cite{Witten:2023qsv,Chandrasekaran:2022cip,Jensen:2023yxy} needs justifications. It would be interesting to construct an explicit microscopic model of the observer like ours, but probably in different contexts. The general framework developed in \cite{Geng:2024dbl} is a good starting point.

\section{Conclusions and Discussions}\label{sec:conclusion}
In this paper, we studied the gravitational subregion algebra in a controlled setup-- the island model, where a well-defined subregion $I$ in a gravitational universe and operators localized within $I$ exist. The subregion is the so-called entanglement island, and the operators localized within this subregion are made consistent with the diffeomorphism constraints, due to the fact that the diffeomorphisms are spontaneously broken. We started with the algebra of these island localized operators and we realize that with a particular element conjoined to this algebra, the algebra becomes unitarily equivalent to the crossed product of the algebra of quantum field theory operators inside the island $I$ with its modular automorphism. Using the classical result from Takesaki \cite{10.1007/BF02392041,takesaki2006tomita}, we imply that the resulting algebra is a Type II$_{\infty}$ von Neumann algebra which is endowed with the notion of trace and entropy. We constructed the trace and studied the entropy. Our work established the result that the emergence of entanglement islands holographically corresponds to the emergence of Type II$_{\infty}$ von Neumann algebras.

The new element we conjoined to the original algebra of island localized operators is a particular mode of the Goldstone vector field. This Goldstone vector field emerges due to the spontaneous breaking of the diffeomorphism symmetry in the island model \cite{Geng:2023ynk,Geng:2023zhq,Geng:2025rov}. Following the notion from \cite{Chandrasekaran:2022cip,Witten:2023qsv,Jensen:1985in}, this new element can be interpreted as an ``observer" and the original island localized operators are made diffeomorphism invariant by being dressed to this ``observer". In contrast to standard quantum systems, the Hamiltonian of this ``observer" is linear in phase space variable. Thus, our model provides an explicit microscopic realization of this so-called ``observer". In fact, a general lesson from our work is that this type of ``observer" is a result of spontaneous breaking of diffeomorphism symmetry, and thus it is ubiquitous in realistic cosmological scenarios \cite{Geng:2024dbl}.

In contrast with earlier work \cite{Chandrasekaran:2022cip,Jensen:2023yxy} in which the observer-dressed operators only obey some of the integrated diffeomorphism constraints , our observer-dressed operators obey all diffeomorphism constraints including the local ones.\footnote{We note that even though our calculation is done only to the leading nontrivial order in $G_{N}$, we expect that our dressed operators could obey the constraints to all orders in gravitational perturbation theory.} Furthermore, our work raises a question for the Type II$_{1}$ constructions in \cite{Chandrasekaran:2022cip,Jensen:2023yxy} where the ``observer" Hamiltonian $\hat{H}_{\text{obs}}$ was projected to be bounded from below. The argument in \cite{Chandrasekaran:2022cip,Jensen:2023yxy} for this projection is that the Hamiltonian of standard quantum systems is bounded from below and thus so is that of the ``observer". Nevertheless, as we discussed, this ``observer" is not a standard quantum system. Thus one shouldn't expect the above lesson to be applied to the ``observer". In fact, in our explicit realization of the ``observer", there is no reason to expect its $\hat{H}_{\text{obs}}$ to be bounded from below. Hence, we believe the projection from \cite{Chandrasekaran:2022cip,Jensen:2023yxy} needs justifications, and this might motivate other realizations of the ``observer" whose Hamiltonian $\hat{H}_{\text{obs}}$ might be indeed bounded from below.
We should emphasize that as in \cite{Jensen:2023yxy}, our work relies on the geometric modular flow conjecture. We refer the readers to \cite{Sorce:2024zme,Caminiti:2025hjq} for recent explorations of the geometric modular flows.

\section*{Acknowledgements}
We are grateful to Yiming Chen, Xi Dong, Steve Giddings, Philipp H\"{o}hn, Kristan Jensen, Andreas Karch, Hong Liu, Henry Maxfield, Suvrat Raju, Lisa Randall and Zhenbin Yang for helpful discussions. HG is supported by the Gravity, Spacetime, and Particle Physics (GRASP) Initiative from Harvard University and a grant from Physics Department at Harvard University. The work of YJ is supported by the U.S Department of Energy ASCR EXPRESS grant, Novel Quantum Algorithms from Fast Classical Transforms, and Northeastern University. The work of J.X. is supported by the U.S. Department of Energy, Office
of Science, Office of High Energy Physics, under Award Number DE-SC0011702. J.X. acknowledges the support by the Graduate Division Dissertation Fellowship and the Physics Department Graduate Fellowship at UCSB.

\appendix
\section{A Lightening Review of the Modular Theory}\label{sec:review1}
In this appendix, we review some basic notation of modular theory relevant for the discussion in the main text, and we also prove some useful properties of Connes cocycle.

We denote by $\mathcal{A}$ the local algebra generated by suitably smeared QFT operators localized in a subregion, acting on a Hilbert space $\mathcal{H}$. Suppose $\Psi \in \mathcal{H}$ is a cyclic and separating state. The Tomita operator $S_\Psi$ is defined by
\begin{equation} \label{eq:def-tomita}
S_\Psi a|\Psi\rangle = a^\dagger |\Psi\rangle, \quad \forall a \in \mathcal{A}.
\end{equation}
The cyclicity of $\Psi$ for $\mathcal{A}$ means that the set $\{a|\Psi\rangle : a \in \mathcal{A}\}$ spans a dense subspace in $\mathcal{H}$, which ensures that $S_\Psi$ is densely defined. The separating property implies that $a|\Psi\rangle = 0$ leads to $a = 0$, thereby removing any ambiguity in the definition Equ.~(\ref{eq:def-tomita})—since a linear operator cannot map a zero vector to non-zero without contradiction. By definition, $S_\Psi$ is anti-linear, meaning it anti-commutes with multiplication by imaginary numbers.

It also follows from the above definition that $S_\Psi$ is generally unbounded. For instance, the vacuum state $\Omega$ in QFT is known to be cyclic and separating. One can construct operators $a$ as suitable superpositions of local operators that nearly annihilate $\Omega$, so that  the norm $\|a|\Omega\rangle\|$ is arbitrarily small. In contrast, $a^\dagger|\Omega\rangle$ can have finite norm, corresponding to an excited state. Hence, $S_\Omega$ following the definition of \eqref{eq:def-tomita} maps a state of arbitrarily small norm to one with finite norm, confirming that it is unbounded.

The modular operator $\Delta_\Psi$ and the modular conjugation $J_\Psi$ associated with $\Psi$ are defined through the unique polar decomposition of $S_\Psi$:
\begin{equation}
S_\Psi = J_\Psi \Delta_\Psi^{1/2},
\end{equation}
where $J_\Psi$ is anti-unitary and satisfies $J_\Psi^2 = 1$, and $\Delta_\Psi = S_\Psi^\dagger S_\Psi$ is a positive-definite, self-adjoint operator. By construction, both operators leave $\Psi$ invariant:
\begin{equation}
J_\Psi |\Psi\rangle = \Delta_\Psi |\Psi\rangle = |\Psi\rangle.
\end{equation}
This structure can be interpreted by viewing $|\Psi\rangle$ as the canonical purification of a thermal equilibrium state in a doubled Hilbert space. The invariance under $\Delta_\Psi$ reflects the triviality of thermal evolution for the equilibrium state, while $J_\Psi$ implements a symmetry between descriptions using operators acting on either factor of the doubled space. The von Neumann algebra formulation generalizes these thermal features to states like $\Psi$ that possess infinite entanglement and cannot be written as elements of a factorized Hilbert space, as typically encountered in QFT on subregions.

Moreover, it follows from \eqref{eq:def-tomita} that $S_\Psi^2 = 1$, which implies the modular conjugation relation:
\begin{equation}
J_\Psi \Delta_\Psi J_\Psi = \Delta_\Psi^{-1}.
\end{equation}
A similar construction applies to the commutant algebra $\mathcal{A}'$. Since a cyclic and separating state for $\mathcal{A}$ is also cyclic and separating for $\mathcal{A}'$, we may define a Tomita operator $S_\Psi'$ for $\mathcal{A}'$:
\begin{equation}
S_\Psi' = J_\Psi \Delta_\Psi^{-1/2}.
\end{equation}
A careful analysis by Tomita-Takesaki theory \cite{Takesaki:1970aki} shows that $J_\Psi$ defines an anti-linear isomorphism between $\mathcal{A}$ and $\mathcal{A}'$, meaning that for every $a \in \mathcal{A}$, the conjugated operator $J_\Psi a J_\Psi$ is an element in $\mathcal{A}'$ and commutes with all elements of $\mathcal{A}$.

We can introduce the modular Hamiltonian $H_{\Psi} = -\log \Delta_{\Psi}$ and write the modular operator as
\begin{equation}
\Delta_{\Psi} = e^{-H_{\Psi}}.
\end{equation}
The modular flow generated by the state $\Psi$ for operators in $\mathcal{A}$ is given by
\begin{equation}
a_s = e^{is H_{\Psi}} a e^{-is H_{\Psi}} = \Delta_{\Psi}^{-is} a \Delta_{\Psi}^{is}, \quad \forall s \in \mathbb{R},\ a \in \mathcal{A},
\end{equation}
which defines a one-parameter family of flowed operators $a_s \in \mathcal{A}$ for any real value of $s$.

In the case where $\mathcal{A}$ is a Type I factor and the Hilbert space $\mathcal{H}$ factorizes as $\mathcal{H} = \mathcal{H}_L \otimes \mathcal{H}_R$, the modular Hamiltonian can be expressed as a difference
\begin{equation}
H_{\Psi} = H_L - H_R,
\end{equation}
with $H_L \in \mathcal{B}(\mathcal{H}_L)$ and $H_R \in \mathcal{B}(\mathcal{H}_R)$ both bounded operators. In this setting, the modular flow becomes an inner automorphism and coincides with time evolution generated by the one-sided Hamiltonian.

For a Type III$_1$ algebra, such a factorization does not exist. The infinite entanglement across the entangling surface prevents any splitting of $H_{\Psi}$ into bounded operators localized on separate sides. As a result, $H_{\Psi}$ no longer belongs to $\mathcal{A}$ nor its commutant $\mathcal{A}'$, and the modular flow becomes an outer automorphism.

The thermal nature of $\Psi$ is reflected in the following KMS-like property of the modular operator:
\begin{equation}
\langle \Psi | a b | \Psi \rangle = \langle \Psi | b \Delta_{\Psi} a | \Psi \rangle, \quad \forall a, b \in \mathcal{A},
\end{equation}
which follows directly from the definition of $\Delta_{\Psi}$. In the Type I case, this identity can be derived from the cyclicity of the trace in a thermal (Gibbs) state:
\begin{equation}
\langle \text{TFD} | ab | \text{TFD} \rangle = \operatorname{Tr}(e^{-H} ab) = \operatorname{Tr}(e^{-H} b e^{-H} a e^{H}) = \langle \text{TFD} | b e^{-(H_L - H_R)} a | \text{TFD} \rangle,
\end{equation}
where both $a$ and $b$ are understood as acting on the left factor, and we have identified $\Psi$ as the thermal field double state $|{\rm TFD}\rangle$ with inverse temperature $\beta=1$.

Given another cyclic and separating state $\Phi$, one can similarly define its modular Hamiltonian $H_{\Phi}$ and study its relation to $H_{\Psi}$. To this end, we define the relative Tomita operator by
\begin{equation}
S_{\Phi|\Psi} a |\Psi\rangle = a^\dagger |\Phi\rangle,
\end{equation}
following the convention of \cite{Chandrasekaran:2022cip,Jensen:2023yxy}. The polar decomposition of $S_{\Phi|\Psi}$ defines the relative modular conjugation $J_{\Phi|\Psi}$ and the relative modular operator $\Delta_{\Phi|\Psi}$. From the definition, it follows that
\begin{equation}
S_{\Psi|\Phi} S_{\Phi|\Psi} = 1,
\end{equation}
which implies the relations
\begin{equation}
J_{\Phi|\Psi}^\dagger = J_{\Psi|\Phi}, \quad J_{\Phi|\Psi} \Delta_{\Phi|\Psi}^{1/2} J_{\Psi|\Phi} = \Delta_{\Psi|\Phi}^{-1/2}.
\end{equation}
A key set of operators that relates the modular flows of different states are the Connes cocycles $u_{\Phi|\Psi}(s)$ and $u_{\Psi|\Phi}'(s)$, defined by
\begin{equation}
\begin{aligned}
u_{\Phi|\Psi}(s) &= \Delta_{\Phi|\Psi}^{is} \Delta_{\Psi}^{-is} = \Delta_{\Phi}^{is} \Delta_{\Psi|\Phi}^{-is}, \\
u_{\Psi|\Phi}'(s) &= \Delta_{\Phi|\Psi}^{-is} \Delta_{\Phi}^{is} = \Delta_{\Psi}^{-is} \Delta_{\Psi|\Phi}^{is},
\end{aligned}
\end{equation}
where in presenting the definition, we already assumed that the two combinations of relative modular operator coincides, which we discuss later.  A crucial property, used in the main text, is that $u_{\Phi|\Psi}(s)$ is a unitary operator belonging to the algebra $\mathcal{A}$, and similarly, $u_{\Psi|\Phi}'(s)$ belongs to the com mutant $\mathcal{A}'$. We will prove this fact in the following:

\paragraph{Basic Properties of Connes Cocycle} We now proceed to show that the Connes cocycle indeed lies in the subregion algebra. The original proof was established in \cite{Connes1973}, and the idea is to consider the modular operator of the subregion algebra with an auxiliary qubit appended.  

Let \( Q, \bar{Q} = \mathbb{C}^2 \) be the Hilbert space of two qubits, and let \( \Omega \) and \( \Psi \) be cyclic and separating states for the local algebras \( \mathcal{A} \) and \( \mathcal{A}' \), respectively. Define the state \( |\Theta\rangle \in \mathcal{H}_1 = \mathcal{H} \otimes Q \otimes \bar{Q} \) as
\begin{equation}
|\Theta\rangle = |\Omega\rangle \otimes |00\rangle + |\Psi\rangle \otimes |11\rangle.
\end{equation}
It is helpful to represent \( \Theta \) as a matrix in \( \mathbb{C}^2 \times \mathbb{C}^2 \), with diagonal components given by \( \Omega \) and \( \Psi \), respectively:
\begin{equation}
|\Theta\rangle = \begin{pmatrix}
|\Omega\rangle & 0 \\
0 & |\Psi\rangle
\end{pmatrix}.
\end{equation}
Operators in the enlarged algebra \( \mathcal{A}_1 = \mathcal{A} \otimes \mathcal{L}^2(Q) \) can be represented as matrix operators:
\begin{equation}
a = \begin{pmatrix}
a_1 & a_2 \\
a_3 & a_4
\end{pmatrix}, \quad a \in \mathcal{A} \otimes \mathcal{L}^2(Q), \quad a_i \in \mathcal{A},
\end{equation}
and similar for $a^\prime\in\mathcal{A}^{\prime}_1= \mathcal{A}^{\prime}\otimes \mathcal{L}^2 (\mathbb{\bar{Q}})$. They act on state $|\Theta\rangle$ from two sides
\begin{equation} \label{eq:action}
a|\Theta\rangle = \begin{pmatrix}
a_1 |\Omega\rangle & a_2 |\Psi\rangle \\
a_3 |\Omega\rangle & a_4 |\Psi\rangle
\end{pmatrix}, \quad a^\prime |\Theta\rangle = \begin{pmatrix}
a^{\prime}_1 |\Omega\rangle & a^{\prime}_2 |\Omega\rangle \\
a^{\prime}_3 |\Psi\rangle & a^{\prime}_4 |\Psi\rangle
\end{pmatrix}\,,
\end{equation}
and therefore commutes with each other. Note that states of the form \eqref{eq:action} spans a dense subspace of \( \mathcal{H}_1 \). Therefore, \( \Theta \) is cyclic and separating for both \( \mathcal{A}_1 \) and \( \mathcal{A}_1' \). 

We now consider the Tomita operator \( S_\Theta \), defined as
\begin{equation}
S_\Theta x |\Theta\rangle = x^\dagger |\Theta\rangle, \quad \forall x \in \mathcal{A}_1.
\end{equation}
Expanding in components, we find
\begin{equation}
S_\Theta a|\Theta\rangle = \begin{pmatrix}
a_1^\dagger |\Omega\rangle & a_3^\dagger |\Psi\rangle \\
a_2^\dagger |\Omega\rangle & a_4^\dagger |\Psi\rangle
\end{pmatrix}
= \begin{pmatrix}
S_\Omega a_1 |\Omega\rangle & S_{\Psi|\Omega} a_3 |\Omega\rangle \\
S_{\Omega|\Psi} a_2 |\Psi\rangle & S_\Psi a_4 |\Psi\rangle
\end{pmatrix}.
\end{equation}
That is, the components of \( S_\Theta \) include both Tomita operators for \( \Omega \) and \( \Psi \), as well as their relative Tomita operators:
\begin{equation}
S_\Theta = S_\Omega \otimes |00\rangle\langle00| + S_\Psi \otimes |11\rangle\langle11| + S_{\Omega|\Psi} \otimes |10\rangle\langle01| + S_{\Psi|\Omega} \otimes |01\rangle\langle10|.
\end{equation}
The corresponding modular operator is then
\begin{equation}
\Delta_\Theta = S_\Theta^\dagger S_\Theta = \Delta_\Omega \otimes |00\rangle\langle00| + \Delta_\Psi \otimes |11\rangle\langle11| + \Delta_{\Psi|\Omega} \otimes |10\rangle\langle10| + \Delta_{\Omega|\Psi} \otimes |01\rangle\langle01|.
\end{equation}
Now consider the modular flow of the operator \( \mathbf{1}_{\mathcal{A}} \otimes |0\rangle\langle1|_Q \). We compute
\begin{equation}
\begin{aligned}
\Delta_\Theta^{it} \left( \mathbf{1}_{\mathcal{A}} \otimes |0\rangle\langle1|_Q \otimes \mathbf{1}_{\mathcal{A}_1'} \right) \Delta_\Theta^{-it}
= & \, \Delta_\Omega^{it} \Delta_{\Psi|\Omega}^{-it} \otimes |00\rangle\langle10| \\
& + \Delta_{\Omega|\Psi}^{it} \Delta_\Psi^{-it} \otimes |01\rangle\langle11|.
\end{aligned}
\end{equation}
Since conjugation by \( \Delta_\Theta^{it} \) defines an automorphism of \( \mathcal{A}_1 \), the result must be of the form \( x \otimes \mathbf{1}_{\mathcal{A}_1'} \) for some \( x \in \mathcal{A}_1 \). This is possible only if
\begin{equation}
\Delta_\Omega^{it} \Delta_{\Psi|\Omega}^{-it} = \Delta_{\Omega|\Psi}^{it} \Delta_\Psi^{-it} \equiv u_{\Omega|\Psi}(t) \in \mathcal{A}.
\end{equation}
This confirms that the Connes cocycle \( u_{\Omega|\Psi}(t) \) indeed lies in the algebra \( \mathcal{A} \).

\bibliographystyle{JHEP}

\bibliography{main}

\end{document}